%
%
%
%
%
%
%
%
\def\standardrisposta{s }\def\reducedrisposta{r }
\def\mplarisposta{mpla }\def\zerorisposta{z }
\def\doublerisposta{d }\def\cartarisposta{e }\def\amsrisposta{y }
\newcount\ingrandimento \newcount\sinnota \newcount\dimnota
\newcount\unoduecol \newdimen\collhsize \newdimen\tothsize
\newdimen\fullhsize \newcount\controllorisposta \sinnota=1
\newskip\infralinea  \global\controllorisposta=0
%
%
%
%
%
\def\risposta{s }
\def\srisposta{e }
\def\arisposta{y }
\ifx\risposta\standardrisposta \ingrandimento=1200
\message {>> This will come out UNREDUCED << }
\dimnota=2 \unoduecol=1 \global\controllorisposta=1 \fi
\ifx\risposta\reducedrisposta \ingrandimento=1095 \dimnota=1
\unoduecol=1  \global\controllorisposta=1
\message {>> This will come out REDUCED << } \fi
\ifx\risposta\doublerisposta \ingrandimento=1000 \dimnota=2
\unoduecol=2   \message {>> You must print this in
LANDSCAPE orientation << } \global\controllorisposta=1 \fi
\ifx\risposta\mplarisposta \ingrandimento=1000 \dimnota=1
\message {>> Mod. Phys. Lett. A format << }
\unoduecol=1 \global\controllorisposta=1 \fi
\ifx\risposta\zerorisposta \ingrandimento=1000 \dimnota=2
\message {>> Zero Magnification format << }
\unoduecol=1 \global\controllorisposta=1 \fi
\ifnum\controllorisposta=0  \ingrandimento=1200
\message {>>> ERROR IN INPUT, I ASSUME STANDARD
UNREDUCED FORMAT <<< }  \dimnota=2 \unoduecol=1 \fi
\magnification=\ingrandimento
%
%
%
%
\newdimen\eucolumnsize \newdimen\eudoublehsize \newdimen\eudoublevsize
\newdimen\uscolumnsize \newdimen\usdoublehsize \newdimen\usdoublevsize
\newdimen\eusinglehsize \newdimen\eusinglevsize \newdimen\ussinglehsize
\newskip\standardbaselineskip \newdimen\ussinglevsize
\newskip\reducedbaselineskip \newskip\doublebaselineskip
\eucolumnsize=12.0truecm    
\eudoublehsize=25.5truecm   
\eudoublevsize=6.5truein    
\uscolumnsize=4.4truein     
\usdoublehsize=9.4truein    
\usdoublevsize=6.8truein    
\eusinglehsize=6.5truein    
\eusinglevsize=24truecm     
\ussinglehsize=6.5truein    
\ussinglevsize=8.9truein    
\standardbaselineskip=16pt plus.2pt  
\reducedbaselineskip=14pt plus.2pt   
\doublebaselineskip=12pt plus.2pt    
%
%
\def\Portoffset{}
\def\Landoffset{}
\ifx\risposta\mplarisposta \def\Portoffset{\hoffset=1.8truecm} \fi
%
%
\def\Landspec{}
\tolerance=10000
\parskip=0pt plus2pt  \leftskip=0pt \rightskip=0pt
%
%
\ifx\risposta\standardrisposta \infralinea=\standardbaselineskip \fi
\ifx\risposta\reducedrisposta  \infralinea=\reducedbaselineskip \fi
\ifx\risposta\doublerisposta   \infralinea=\doublebaselineskip \fi
\ifx\risposta\mplarisposta     \infralinea=13pt \fi
\ifx\risposta\zerorisposta     \infralinea=12pt plus.2pt\fi
\ifnum\controllorisposta=0    \infralinea=\standardbaselineskip \fi
\ifx\risposta\doublerisposta   \Landoffset \else \Portoffset \fi
\ifx\risposta\doublerisposta \ifx\srisposta\cartarisposta
\tothsize=\eudoublehsize \collhsize=\eucolumnsize
\vsize=\eudoublevsize  \else  \tothsize=\usdoublehsize
\collhsize=\uscolumnsize \vsize=\usdoublevsize \fi \else
\ifx\srisposta\cartarisposta \tothsize=\eusinglehsize
\vsize=\eusinglevsize \else  \tothsize=\ussinglehsize
\vsize=\ussinglevsize \fi \collhsize=4.4truein \fi
\ifx\risposta\mplarisposta \tothsize=5.0truein
\vsize=7.8truein \collhsize=4.4truein \fi
%
%
%
%
\newcount\contaeuler \newcount\contacyrill \newcount\contaams
\font\ninerm=cmr9  \font\eightrm=cmr8  \font\sixrm=cmr6
\font\ninei=cmmi9  \font\eighti=cmmi8  \font\sixi=cmmi6
\font\ninesy=cmsy9  \font\eightsy=cmsy8  \font\sixsy=cmsy6
\font\ninebf=cmbx9  \font\eightbf=cmbx8  \font\sixbf=cmbx6
\font\ninett=cmtt9  \font\eighttt=cmtt8  \font\nineit=cmti9
\font\eightit=cmti8 \font\ninesl=cmsl9  \font\eightsl=cmsl8
\skewchar\ninei='177 \skewchar\eighti='177 \skewchar\sixi='177
\skewchar\ninesy='60 \skewchar\eightsy='60 \skewchar\sixsy='60
\hyphenchar\ninett=-1 \hyphenchar\eighttt=-1 \hyphenchar\tentt=-1
\def\bfmath{\cmmib}                 
\font\tencmmib=cmmib10  \newfam\cmmibfam  \skewchar\tencmmib='177
\font\tencmbsy=cmbsy10  \newfam\cmbsyfam  \skewchar\tencmbsy='60
\def\scaps{\cmcsc}                 
\font\tencmcsc=cmcsc10  \newfam\cmcscfam
\ifnum\ingrandimento=1095

\font\capsone=cmcsc10 at 10.95pt 

\else

\font\capsone=cmcsc10 at 12pt 
\fi

\def\ttaarr{\bf}		
\def\ppaarr{\sl}		

%
%
%
\newfam\eufmfam \newfam\msamfam \newfam\msbmfam \newfam\eufbfam
\def\Loadeulerfonts{\global\contaeuler=1 \ifx\arisposta\amsrisposta
\font\teneufm=eufm10              
\font\eighteufm=eufm8 \font\nineeufm=eufm9 \font\sixeufm=eufm6
\font\seveneufm=eufm7  \font\fiveeufm=eufm5
\font\teneufb=eufb10              
\font\eighteufb=eufb8 \font\nineeufb=eufb9 \font\sixeufb=eufb6
\font\seveneufb=eufb7  \font\fiveeufb=eufb5
\font\teneurm=eurm10              
\font\eighteurm=eurm8 \font\nineeurm=eurm9
\font\teneurb=eurb10              
\font\eighteurb=eurb8 \font\nineeurb=eurb9
\font\teneusm=eusm10              
\font\eighteusm=eusm8 \font\nineeusm=eusm9
\font\teneusb=eusb10              
\font\eighteusb=eusb8 \font\nineeusb=eusb9
\else \def\eufm{\tt} \def\eufb{\tt} \def\eurm{\tt} \def\eurb{\tt}
\def\eusm{\tt} \def\eusb{\tt}    \fi}
\def\loadeuler{\Loadeulerfonts\tenpoint}
\def\loadamsmath{\global\contaams=1 \ifx\arisposta\amsrisposta
\font\tenmsam=msam10 \font\ninemsam=msam9 \font\eightmsam=msam8
\font\sevenmsam=msam7 \font\sixmsam=msam6 \font\fivemsam=msam5
\font\tenmsbm=msbm10 \font\ninemsbm=msbm9 \font\eightmsbm=msbm8
\font\sevenmsbm=msbm7 \font\sixmsbm=msbm6 \font\fivemsbm=msbm5
\else \def\msbm{\bf} \fi \def\Bbb{\msbm} \def\symbl{\msam} \tenpoint}
\def\loadcyrill{\global\contacyrill=1 \ifx\arisposta\amsrisposta
\font\tenwncyr=wncyr10 \font\ninewncyr=wncyr9 \font\eightwncyr=wncyr8
\font\tenwncyb=wncyr10 \font\ninewncyb=wncyr9 \font\eightwncyb=wncyr8
\font\tenwncyi=wncyr10 \font\ninewncyi=wncyr9 \font\eightwncyi=wncyr8
\else \def\cyrill{\sl} \def\cyrilb{\sl} \def\cyrili{\sl} \fi\tenpoint}
\ifx\arisposta\amsrisposta
\font\sevenex=cmex7               
\font\eightex=cmex8  \font\nineex=cmex9
\font\ninecmmib=cmmib9   \font\eightcmmib=cmmib8
\font\sevencmmib=cmmib7 \font\sixcmmib=cmmib6
\font\fivecmmib=cmmib5   \skewchar\ninecmmib='177
\skewchar\eightcmmib='177  \skewchar\sevencmmib='177
\skewchar\sixcmmib='177   \skewchar\fivecmmib='177
\font\ninecmbsy=cmbsy9    \font\eightcmbsy=cmbsy8
\font\sevencmbsy=cmbsy7  \font\sixcmbsy=cmbsy6
\font\fivecmbsy=cmbsy5   \skewchar\ninecmbsy='60
\skewchar\eightcmbsy='60  \skewchar\sevencmbsy='60
\skewchar\sixcmbsy='60    \skewchar\fivecmbsy='60
\font\ninecmcsc=cmcsc9    \font\eightcmcsc=cmcsc8     \else
\def\cmmib{\fam\cmmibfam\tencmmib}\textfont\cmmibfam=\tencmmib
\scriptfont\cmmibfam=\tencmmib \scriptscriptfont\cmmibfam=\tencmmib
\def\cmbsy{\fam\cmbsyfam\tencmbsy} \textfont\cmbsyfam=\tencmbsy
\scriptfont\cmbsyfam=\tencmbsy \scriptscriptfont\cmbsyfam=\tencmbsy
\scriptfont\cmcscfam=\tencmcsc \scriptscriptfont\cmcscfam=\tencmcsc
\def\cmcsc{\fam\cmcscfam\tencmcsc} \textfont\cmcscfam=\tencmcsc \fi
\catcode`@=11
\newskip\ttglue
\gdef\tenpoint{\def\rm{\fam0\tenrm}
  \textfont0=\tenrm \scriptfont0=\sevenrm \scriptscriptfont0=\fiverm
  \textfont1=\teni \scriptfont1=\seveni \scriptscriptfont1=\fivei
  \textfont2=\tensy \scriptfont2=\sevensy \scriptscriptfont2=\fivesy
  \textfont3=\tenex \scriptfont3=\tenex \scriptscriptfont3=\tenex
  \def\mcal{\fam2 \tensy}  \def\mmit{\fam1 \teni}
  \textfont\itfam=\tenit \def\it{\fam\itfam\tenit}
  \textfont\slfam=\tensl \def\sl{\fam\slfam\tensl}
  \textfont\ttfam=\tentt \scriptfont\ttfam=\eighttt
  \scriptscriptfont\ttfam=\eighttt  \def\tt{\fam\ttfam\tentt}
  \textfont\bffam=\tenbf \scriptfont\bffam=\sevenbf
  \scriptscriptfont\bffam=\fivebf \def\bf{\fam\bffam\tenbf}
     \ifx\arisposta\amsrisposta    \ifnum\contaeuler=1
  \textfont\eufmfam=\teneufm \scriptfont\eufmfam=\seveneufm
  \scriptscriptfont\eufmfam=\fiveeufm \def\eufm{\fam\eufmfam\teneufm}
  \textfont\eufbfam=\teneufb \scriptfont\eufbfam=\seveneufb
  \scriptscriptfont\eufbfam=\fiveeufb \def\eufb{\fam\eufbfam\teneufb}
  \def\eurm{\teneurm} \def\eurb{\teneurb} \def\eusm{\teneusm}
  \def\eusb{\teneusb}    \fi    \ifnum\contaams=1
  \textfont\msamfam=\tenmsam \scriptfont\msamfam=\sevenmsam
  \scriptscriptfont\msamfam=\fivemsam \def\msam{\fam\msamfam\tenmsam}
  \textfont\msbmfam=\tenmsbm \scriptfont\msbmfam=\sevenmsbm
  \scriptscriptfont\msbmfam=\fivemsbm \def\msbm{\fam\msbmfam\tenmsbm}
     \fi      \ifnum\contacyrill=1     \def\cyrill{\tenwncyr}
  \def\cyrilb{\tenwncyb}  \def\cyrili{\tenwncyi}         \fi
  \textfont3=\tenex \scriptfont3=\sevenex \scriptscriptfont3=\sevenex
  \def\cmmib{\fam\cmmibfam\tencmmib} \scriptfont\cmmibfam=\sevencmmib
  \textfont\cmmibfam=\tencmmib  \scriptscriptfont\cmmibfam=\fivecmmib
  \def\cmbsy{\fam\cmbsyfam\tencmbsy} \scriptfont\cmbsyfam=\sevencmbsy
  \textfont\cmbsyfam=\tencmbsy  \scriptscriptfont\cmbsyfam=\fivecmbsy
  \def\cmcsc{\fam\cmcscfam\tencmcsc} \scriptfont\cmcscfam=\eightcmcsc
  \textfont\cmcscfam=\tencmcsc \scriptscriptfont\cmcscfam=\eightcmcsc
     \fi            \tt \ttglue=.5em plus.25em minus.15em
  \normalbaselineskip=12pt
  \setbox\strutbox=\hbox{\vrule height8.5pt depth3.5pt width0pt}
  \let\sc=\eightrm \let\big=\tenbig   \normalbaselines
  \baselineskip=\infralinea  \rm}
\gdef\ninepoint{\def\rm{\fam0\ninerm}
  \textfont0=\ninerm \scriptfont0=\sixrm \scriptscriptfont0=\fiverm
  \textfont1=\ninei \scriptfont1=\sixi \scriptscriptfont1=\fivei
  \textfont2=\ninesy \scriptfont2=\sixsy \scriptscriptfont2=\fivesy
  \textfont3=\tenex \scriptfont3=\tenex \scriptscriptfont3=\tenex
  \def\mcal{\fam2 \ninesy}  \def\mmit{\fam1 \ninei}
  \textfont\itfam=\nineit \def\it{\fam\itfam\nineit}
  \textfont\slfam=\ninesl \def\sl{\fam\slfam\ninesl}
  \textfont\ttfam=\ninett \scriptfont\ttfam=\eighttt
  \scriptscriptfont\ttfam=\eighttt \def\tt{\fam\ttfam\ninett}
  \textfont\bffam=\ninebf \scriptfont\bffam=\sixbf
  \scriptscriptfont\bffam=\fivebf \def\bf{\fam\bffam\ninebf}
     \ifx\arisposta\amsrisposta  \ifnum\contaeuler=1
  \textfont\eufmfam=\nineeufm \scriptfont\eufmfam=\sixeufm
  \scriptscriptfont\eufmfam=\fiveeufm \def\eufm{\fam\eufmfam\nineeufm}
  \textfont\eufbfam=\nineeufb \scriptfont\eufbfam=\sixeufb
  \scriptscriptfont\eufbfam=\fiveeufb \def\eufb{\fam\eufbfam\nineeufb}
  \def\eurm{\nineeurm} \def\eurb{\nineeurb} \def\eusm{\nineeusm}
  \def\eusb{\nineeusb}     \fi   \ifnum\contaams=1
  \textfont\msamfam=\ninemsam \scriptfont\msamfam=\sixmsam
  \scriptscriptfont\msamfam=\fivemsam \def\msam{\fam\msamfam\ninemsam}
  \textfont\msbmfam=\ninemsbm \scriptfont\msbmfam=\sixmsbm
  \scriptscriptfont\msbmfam=\fivemsbm \def\msbm{\fam\msbmfam\ninemsbm}
     \fi       \ifnum\contacyrill=1     \def\cyrill{\ninewncyr}
  \def\cyrilb{\ninewncyb}  \def\cyrili{\ninewncyi}         \fi
  \textfont3=\nineex \scriptfont3=\sevenex \scriptscriptfont3=\sevenex
  \def\cmmib{\fam\cmmibfam\ninecmmib}  \textfont\cmmibfam=\ninecmmib
  \scriptfont\cmmibfam=\sixcmmib \scriptscriptfont\cmmibfam=\fivecmmib
  \def\cmbsy{\fam\cmbsyfam\ninecmbsy}  \textfont\cmbsyfam=\ninecmbsy
  \scriptfont\cmbsyfam=\sixcmbsy \scriptscriptfont\cmbsyfam=\fivecmbsy
  \def\cmcsc{\fam\cmcscfam\ninecmcsc} \scriptfont\cmcscfam=\eightcmcsc
  \textfont\cmcscfam=\ninecmcsc \scriptscriptfont\cmcscfam=\eightcmcsc
     \fi            \tt \ttglue=.5em plus.25em minus.15em
  \normalbaselineskip=11pt
  \setbox\strutbox=\hbox{\vrule height8pt depth3pt width0pt}
  \let\sc=\sevenrm \let\big=\ninebig \normalbaselines\rm}
\gdef\eightpoint{\def\rm{\fam0\eightrm}
  \textfont0=\eightrm \scriptfont0=\sixrm \scriptscriptfont0=\fiverm
  \textfont1=\eighti \scriptfont1=\sixi \scriptscriptfont1=\fivei
  \textfont2=\eightsy \scriptfont2=\sixsy \scriptscriptfont2=\fivesy
  \textfont3=\tenex \scriptfont3=\tenex \scriptscriptfont3=\tenex
  \def\mcal{\fam2 \eightsy}  \def\mmit{\fam1 \eighti}
  \textfont\itfam=\eightit \def\it{\fam\itfam\eightit}
  \textfont\slfam=\eightsl \def\sl{\fam\slfam\eightsl}
  \textfont\ttfam=\eighttt \scriptfont\ttfam=\eighttt
  \scriptscriptfont\ttfam=\eighttt \def\tt{\fam\ttfam\eighttt}
  \textfont\bffam=\eightbf \scriptfont\bffam=\sixbf
  \scriptscriptfont\bffam=\fivebf \def\bf{\fam\bffam\eightbf}
     \ifx\arisposta\amsrisposta   \ifnum\contaeuler=1
  \textfont\eufmfam=\eighteufm \scriptfont\eufmfam=\sixeufm
  \scriptscriptfont\eufmfam=\fiveeufm \def\eufm{\fam\eufmfam\eighteufm}
  \textfont\eufbfam=\eighteufb \scriptfont\eufbfam=\sixeufb
  \scriptscriptfont\eufbfam=\fiveeufb \def\eufb{\fam\eufbfam\eighteufb}
  \def\eurm{\eighteurm} \def\eurb{\eighteurb} \def\eusm{\eighteusm}
  \def\eusb{\eighteusb}       \fi    \ifnum\contaams=1
  \textfont\msamfam=\eightmsam \scriptfont\msamfam=\sixmsam
  \scriptscriptfont\msamfam=\fivemsam \def\msam{\fam\msamfam\eightmsam}
  \textfont\msbmfam=\eightmsbm \scriptfont\msbmfam=\sixmsbm
  \scriptscriptfont\msbmfam=\fivemsbm \def\msbm{\fam\msbmfam\eightmsbm}
     \fi       \ifnum\contacyrill=1     \def\cyrill{\eightwncyr}
  \def\cyrilb{\eightwncyb}  \def\cyrili{\eightwncyi}         \fi
  \textfont3=\eightex \scriptfont3=\sevenex \scriptscriptfont3=\sevenex
  \def\cmmib{\fam\cmmibfam\eightcmmib}  \textfont\cmmibfam=\eightcmmib
  \scriptfont\cmmibfam=\sixcmmib \scriptscriptfont\cmmibfam=\fivecmmib
  \def\cmbsy{\fam\cmbsyfam\eightcmbsy}  \textfont\cmbsyfam=\eightcmbsy
  \scriptfont\cmbsyfam=\sixcmbsy \scriptscriptfont\cmbsyfam=\fivecmbsy
  \def\cmcsc{\fam\cmcscfam\eightcmcsc} \scriptfont\cmcscfam=\eightcmcsc
  \textfont\cmcscfam=\eightcmcsc \scriptscriptfont\cmcscfam=\eightcmcsc
     \fi             \tt \ttglue=.5em plus.25em minus.15em
  \normalbaselineskip=9pt
  \setbox\strutbox=\hbox{\vrule height7pt depth2pt width0pt}
  \let\sc=\sixrm \let\big=\eightbig \normalbaselines\rm }
\gdef\tenbig#1{{\hbox{$\left#1\vbox to8.5pt{}\right.\n@space$}}}
\gdef\ninebig#1{{\hbox{$\textfont0=\tenrm\textfont2=\tensy
   \left#1\vbox to7.25pt{}\right.\n@space$}}}
\gdef\eightbig#1{{\hbox{$\textfont0=\ninerm\textfont2=\ninesy
   \left#1\vbox to6.5pt{}\right.\n@space$}}}
\def\alternativefont#1#2{\ifx\arisposta\amsrisposta \relax \else
\xdef#1{#2} \fi}
\global\contaeuler=0 \global\contacyrill=0 \global\contaams=0
%
%
%
%
\newbox\fotlinebb \newbox\hedlinebb \newbox\leftcolumn
\gdef\makeheadline{\vbox to 0pt{\vskip-22.5pt
     \fullline{\vbox to8.5pt{}\the\headline}\vss}\nointerlineskip}
\gdef\makehedlinebb{\vbox to 0pt{\vskip-22.5pt
     \fullline{\vbox to8.5pt{}\copy\hedlinebb\hfil
     \line{\hfill\the\headline\hfill}}\vss} \nointerlineskip}
\gdef\makefootline{\baselineskip=24pt \fullline{\the\footline}}
\gdef\makefotlinebb{\baselineskip=24pt
    \fullline{\copy\fotlinebb\hfil\line{\hfill\the\footline\hfill}}}
\gdef\doubleformat{\shipout\vbox{\Landspec\makehedlinebb
     \fullline{\box\leftcolumn\hfil\columnbox}\makefotlinebb}
     \advancepageno}
\gdef\columnbox{\leftline{\pagebody}}
\gdef\line#1{\hbox to\hsize{\hskip\leftskip#1\hskip\rightskip}}
\gdef\fullline#1{\hbox to\fullhsize{\hskip\leftskip{#1}%
\hskip\rightskip}}
\gdef\footnote#1{\let\@sf=\empty
         \ifhmode\edef\#sf{\spacefactor=\the\spacefactor}\/\fi
         #1\@sf\vfootnote{#1}}
\gdef\vfootnote#1{\insert\footins\bgroup
         \ifnum\dimnota=1  \eightpoint\fi
         \ifnum\dimnota=2  \ninepoint\fi
         \ifnum\dimnota=0  \tenpoint\fi
         \interlinepenalty=\interfootnotelinepenalty
         \splittopskip=\ht\strutbox
         \splitmaxdepth=\dp\strutbox \floatingpenalty=20000
         \leftskip=\oldssposta \rightskip=\olddsposta
         \spaceskip=0pt \xspaceskip=0pt
         \ifnum\sinnota=0   \textindent{#1}\fi
         \ifnum\sinnota=1   \item{#1}\fi
         \footstrut\futurelet\next\fo@t}
\gdef\fo@t{\ifcat\bgroup\noexpand\next \let\next\f@@t
             \else\let\next\f@t\fi \next}
\gdef\f@@t{\bgroup\aftergroup\@foot\let\next}
\gdef\f@t#1{#1\@foot} \gdef\@foot{\strut\egroup}
\gdef\footstrut{\vbox to\splittopskip{}}
\skip\footins=\bigskipamount
\count\footins=1000  \dimen\footins=8in
\catcode`@=12
\tenpoint
\ifnum\unoduecol=1 \hsize=\tothsize   \fullhsize=\tothsize \fi
\ifnum\unoduecol=2 \hsize=\collhsize  \fullhsize=\tothsize \fi
\global\let\lrcol=L      \ifnum\unoduecol=1
\output{\plainoutput{\ifnum\tipbnota=2 \clearnmbnota\fi}} \fi
\ifnum\unoduecol=2 \output{\if L\lrcol
     \global\setbox\leftcolumn=\columnbox
     \global\setbox\fotlinebb=\line{\hfill\the\footline\hfill}
     \global\setbox\hedlinebb=\line{\hfill\the\headline\hfill}
     \advancepageno  \global\let\lrcol=R
     \else  \doubleformat \global\let\lrcol=L \fi
     \ifnum\outputpenalty>-20000 \else\dosupereject\fi
     \ifnum\tipbnota=2\clearnmbnota\fi }\fi
\def\ifdoublepage{\ifnum\unoduecol=2 }
\gdef\yespagenumbers{\footline={\hss\tenrm\folio\hss}}
\gdef\ciao{ \ifnum\fdefcontre=1 \endfdef\fi
     \par\vfill\supereject \ifnum\unoduecol=2
     \if R\lrcol  \headline={}\nopagenumbers\null\vfill\eject
     \fi\fi \end}

\newskip\olddsposta \newskip\oldssposta
\global\oldssposta=\leftskip \global\olddsposta=\rightskip

\def\filldots{\leaders\hbox to 1em{\hss.\hss}\hfill}
\def\inquadrb#1 {\vbox {\hrule  \hbox{\vrule \vbox {\vskip .2cm
    \hbox {\ #1\ } \vskip .2cm } \vrule  }  \hrule} }
 \def\newline{\hfil\break}
\def\jump{\vskip\baselineskip} \newskip\iinnffrr
\def\sjump{\iinnffrr=\baselineskip
          \divide\iinnffrr by 2 \vskip\iinnffrr}
\def\bjump{\vskip\baselineskip \vskip\baselineskip}
\newcount\nmbnota  \def\clearnmbnota{\global\nmbnota=0}
\newcount\tipbnota \def\letterfootnote{\global\tipbnota=1}

\def\note#1{\global\advance\nmbnota by 1 \ifnum\tipbnota=1
    \footnote{$^{\rm\nttlett}$}{#1} \else {\ifnum\tipbnota=2
    \footnote{$^{\nttsymb}$}{#1}
    \else\footnote{$^{\the\nmbnota}$}{#1}\fi}\fi}
\def\nttlett{\ifcase\nmbnota \or a\or b\or c\or d\or e\or f\or
g\or h\or i\or j\or k\or l\or m\or n\or o\or p\or q\or r\or
s\or t\or u\or v\or w\or y\or x\or z\fi}
\def\nttsymb{\ifcase\nmbnota \or\dag\or\sharp\or\ddag\or\star\or
\natural\or\flat\or\clubsuit\or\diamondsuit\or\heartsuit
\or\spadesuit\fi}   \clearnmbnota
\def\numberfootnote{\global\tipbnota=0} \numberfootnote
\def\setnote#1{\expandafter\xdef\csname#1\endcsname{
\ifnum\tipbnota=1 {\rm\nttlett} \else {\ifnum\tipbnota=2
{\nttsymb} \else \the\nmbnota\fi}\fi} }
\newcount\nbmfig  \def\clearnbmfig{\global\nbmfig=0}
\gdef\figure{\global\advance\nbmfig by 1
      {\rm fig. \the\nbmfig}}   \clearnbmfig
\def\setfig#1{\expandafter\xdef\csname#1\endcsname{fig. \the\nbmfig}}
 \def\endformula{\eqno\numero $$}
 \def\efr{\endformula}
\newcount\frmcount \def\clearfrmcount{\global\frmcount=0}
\def\numero{\global\advance\frmcount by 1   \ifnum\indappcount=0
  {\ifnum\cpcount <1 {\hbox{\rm (\the\frmcount )}}  \else
  {\hbox{\rm (\the\cpcount .\the\frmcount )}} \fi}  \else
  {\hbox{\rm (\applett .\the\frmcount )}} \fi}
\def\nameformula#1{\global\advance\frmcount by 1%
\ifnum\draftnum=0  {\ifnum\indappcount=0%
{\ifnum\cpcount<1\xdef\spzzttrra{(\the\frmcount )}%
\else\xdef\spzzttrra{(\the\cpcount .\the\frmcount )}\fi}%
\else\xdef\spzzttrra{(\applett .\the\frmcount )}\fi}%
\else\xdef\spzzttrra{(#1)}\fi%
\expandafter\xdef\csname#1\endcsname{\spzzttrra}
\eqno \hbox{\rm\spzzttrra} $$}
\def\nfr{\nameformula}    
\def\nameali#1{\global\advance\frmcount by 1%
\ifnum\draftnum=0  {\ifnum\indappcount=0%
{\ifnum\cpcount<1\xdef\spzzttrra{(\the\frmcount )}%
\else\xdef\spzzttrra{(\the\cpcount .\the\frmcount )}\fi}%
\else\xdef\spzzttrra{(\applett .\the\frmcount )}\fi}%
\else\xdef\spzzttrra{(#1)}\fi%
\expandafter\xdef\csname#1\endcsname{\spzzttrra}
  \hbox{\rm\spzzttrra} }      \clearfrmcount
\newcount\cpcount \def\clearcpcount{\global\cpcount=0}
\newcount\subcpcount \def\clearsubcpcount{\global\subcpcount=0}
\newcount\appcount \def\clearappcount{\global\appcount=0}
\newcount\indappcount \def\clearindappcount{\indappcount=0}
\newcount\sottoparcount 

\def\applett{\ifcase\appcount  \or {A}\or {B}\or {C}\or
{D}\or {E}\or {F}\or {G}\or {H}\or {I}\or {J}\or {K}\or {L}\or
{M}\or {N}\or {O}\or {P}\or {Q}\or {R}\or {S}\or {T}\or {U}\or
{V}\or {W}\or {X}\or {Y}\or {Z}\fi    \ifnum\appcount<0
\immediate\write16 {Panda ERROR - Appendix: counter "appcount"
out of range}\fi  \ifnum\appcount>26  \immediate\write16 {Panda
ERROR - Appendix: counter "appcount" out of range}\fi}
\clearappcount  \clearindappcount \newcount\connttrre
\def\clearconnttrre{\global\connttrre=0} \newcount\countref
\def\clearcountref{\global\countref=0} \clearcountref
\def\chapter#1{\global\advance\cpcount by 1 \clearfrmcount
                 \goodbreak\null\vbox{\jump\nobreak
                 \clearsubcpcount\clearindappcount
                 \itemitem{\ttaarr\the\cpcount .\qquad}{\ttaarr #1}
                 \par\nobreak\jump\sjump}\nobreak}
\def\section#1{\global\advance\subcpcount by 1 \goodbreak\null
               \vbox{\sjump\nobreak\ifnum\indappcount=0
                 {\ifnum\cpcount=0 {\itemitem{\ppaarr
               .\the\subcpcount\quad\enskip\ }{\ppaarr #1}\par} \else
                 {\itemitem{\ppaarr\the\cpcount .\the\subcpcount\quad
                  \enskip\ }{\ppaarr #1} \par}  \fi}
                \else{\itemitem{\ppaarr\applett .\the\subcpcount\quad
                 \enskip\ }{\ppaarr #1}\par}\fi\nobreak\jump}\nobreak}
\clearsubcpcount
\def\appendix#1{\global\advance\appcount by 1 \clearfrmcount
                  \goodbreak\null\vbox{\jump\nobreak
                  \global\advance\indappcount by 1 \clearsubcpcount
          \itemitem{ }{\hskip-40pt\ttaarr Appendix\ #1}
             \nobreak\jump\sjump}\nobreak}
\clearappcount \clearindappcount
\def\references{\goodbreak\null\vbox{\jump\nobreak
   \itemitem{}{\ttaarr References} \nobreak\jump\sjump}\nobreak}

\clearcpcount\clearcountref

\def\setchap#1{\ifnum\indappcount=0{\ifnum\subcpcount=0%
\xdef\spzzttrra{\the\cpcount}%
\else\xdef\spzzttrra{\the\cpcount .\the\subcpcount}\fi}
\else{\ifnum\subcpcount=0 \xdef\spzzttrra{\applett}%
\else\xdef\spzzttrra{\applett .\the\subcpcount}\fi}\fi
\expandafter\xdef\csname#1\endcsname{\spzzttrra}}
\newcount\draftnum \newcount\ppora   \newcount\ppminuti
\global\ppora=\time   \global\ppminuti=\time
\global\divide\ppora by 60  \draftnum=\ppora
\multiply\draftnum by 60    \global\advance\ppminuti by -\draftnum
\def\droggi{\number\day /\number\month /\number\year\ \the\ppora
:\the\ppminuti}     \global\draftnum=0
\def\draftcomment#1{\ifnum\draftnum=0 \relax \else
{\ {\bf ***}\ #1\ {\bf ***}\ }\fi} 
%
%
\catcode`@=11
\gdef\Ref#1{\expandafter\ifx\csname @rrxx@#1\endcsname\relax%
{\global\advance\countref by 1    \ifnum\countref>200
\immediate\write16 {Panda ERROR - Ref: maximum number of references
exceeded}  \expandafter\xdef\csname @rrxx@#1\endcsname{0}\else
\expandafter\xdef\csname @rrxx@#1\endcsname{\the\countref}\fi}\fi
\ifnum\draftnum=0 \csname @rrxx@#1\endcsname \else#1\fi}
\gdef\beginref{\ifnum\draftnum=0  \gdef\Rref{\fairef}
\gdef\endref{\scriviref} \else\relax\fi
\ifx\risposta\mplarisposta \ninepoint \fi
\parskip 2pt plus.2pt \baselineskip=12pt}
\def\Reflab#1{[#1]} \gdef\Rref#1#2{\item{\Reflab{#1}}{#2}}
\gdef\endref{\relax}  \newcount\conttemp
\gdef\fairef#1#2{\expandafter\ifx\csname @rrxx@#1\endcsname\relax
{\global\conttemp=0 \immediate\write16 {Panda ERROR - Ref: reference
[#1] undefined}} \else
{\global\conttemp=\csname @rrxx@#1\endcsname } \fi
\global\advance\conttemp by 50  \global\setbox\conttemp=\hbox{#2} }
\gdef\scriviref{\clearconnttrre\conttemp=50
\loop\ifnum\connttrre<\countref \advance\conttemp by 1
\advance\connttrre by 1
\item{\Reflab{\the\connttrre}}{\unhcopy\conttemp} \repeat}
\clearcountref \clearconnttrre
\catcode`@=12
\ifx\risposta\mplarisposta \def\Reflab#1{#1.} \letterfootnote \fi

\def\slashchar#1{\setbox0=\hbox{$#1$} \dimen0=\wd0
     \setbox1=\hbox{/} \dimen1=\wd1 \ifdim\dimen0>\dimen1
      \rlap{\hbox to \dimen0{\hfil/\hfil}} #1 \else
      \rlap{\hbox to \dimen1{\hfil$#1$\hfil}} / \fi}
\ifx\oldchi\undefined \let\oldchi=\chi
  \def\cchi{{\raise 1pt\hbox{$\oldchi$}}} \let\chi=\cchi \fi

\def\frac#1#2{{\textstyle{#1 \over #2}}}

\def\half{\ifinner {\scriptstyle {1 \over 2}}\else {1 \over 2} \fi}

\def\simge{\rlap{\raise 2pt \hbox{$>$}}{\lower 2pt \hbox{$\sim$}}}
\def\simle{\rlap{\raise 2pt \hbox{$<$}}{\lower 2pt \hbox{$\sim$}}}

\def\vbig#1#2{{\vbigd@men=#2\divide\vbigd@men by 2%
\hbox{$\left#1\vbox to \vbigd@men{}\right.\n@space$}}}

%
%
\newcount\fdefcontre \newcount\fdefcount \newcount\indcount
\newread\filefdef  \newread\fileftmp  \newwrite\filefdef
\newwrite\fileftmp     \def\strip#1*.A {#1}
\def\futuredef#1{\beginfdef
\expandafter\ifx\csname#1\endcsname\relax%
{\immediate\write\fileftmp {#1*.A}
\immediate\write16 {Panda Warning - fdef: macro "#1" on page
\the\pageno \space undefined}
\ifnum\draftnum=0 \expandafter\xdef\csname#1\endcsname{(?)}
\else \expandafter\xdef\csname#1\endcsname{(#1)} \fi
\global\advance\fdefcount by 1}\fi   \csname#1\endcsname}

\def\beginfdef{\ifnum\fdefcontre=0
\immediate\openin\filefdef \jobname.fdef
\immediate\openout\fileftmp \jobname.ftmp
\global\fdefcontre=1  \ifeof\filefdef \immediate\write16 {Panda
WARNING - fdef: file \jobname.fdef not found, run TeX again}
\else \immediate\read\filefdef to\spzzttrra
\global\advance\fdefcount by \spzzttrra
\indcount=0      \loop\ifnum\indcount<\fdefcount
\advance\indcount by 1   \immediate\read\filefdef to\spezttrra
\immediate\read\filefdef to\sppzttrra
\edef\spzzttrra{\expandafter\strip\spezttrra}
\immediate\write\fileftmp {\spzzttrra *.A}
\expandafter\xdef\csname\spzzttrra\endcsname{\sppzttrra}
\repeat \fi \immediate\closein\filefdef \fi}
\def\endfdef{\immediate\closeout\fileftmp   \ifnum\fdefcount>0
\immediate\openin\fileftmp \jobname.ftmp
\immediate\openout\filefdef \jobname.fdef
\immediate\write\filefdef {\the\fdefcount}   \indcount=0
\loop\ifnum\indcount<\fdefcount    \advance\indcount by 1
\immediate\read\fileftmp to\spezttrra
\edef\spzzttrra{\expandafter\strip\spezttrra}
\immediate\write\filefdef{\spzzttrra *.A}
\edef\spezttrra{\string{\csname\spzzttrra\endcsname\string}}
\iwritel\filefdef{\spezttrra}
\repeat  \immediate\closein\fileftmp \immediate\closeout\filefdef
\immediate\write16 {Panda Warning - fdef: Label(s) may have changed,
re-run TeX to get them right}\fi}
\def\iwritel#1#2{\newlinechar=-1
{\newlinechar=`\ \immediate\write#1{#2}}\newlinechar=-1}
\global\fdefcontre=0 \global\fdefcount=0 \global\indcount=0
%
%
\null
%
%
%
%

\input psfig

%
\loadamsmath
\loadeuler
\mathchardef\bbalpha="710B
\mathchardef\bbbeta="710C
\mathchardef\bbgamma="710D
\mathchardef\bbdelta="710E
\mathchardef\bblambda="7115
\mathchardef\bbxi="7118
\mathchardef\bbpsi="7112
\mathchardef\bbrho="7116
\mathchardef\bbomega="7121
\mathchardef\sdir="2D6E
\mathchardef\dirs="2D6F

\def\bal{{\bfmath\bbalpha}}
\def\bb{{\bfmath\bbbeta}}
\def\bgamma{{\bfmath\bbgamma}}
\def\bd{{\bfmath\bbdelta}}

\def\ba{{\bfmath a}}

\def\bq{{\bfmath q}}
\def\bg{{\bfmath g}}
\def\bH{{\bfmath H}}

\pageno=0\baselineskip=14pt\parskip10pt
\nopagenumbers{
\line{\hfill SWAT/144}
\line{\hfill\tt hep-th/9705041}
\line{\hfill May 1997}
\ifdoublepage \bjump\bjump\bjump\bjump\else\vfill\fi
\centerline{\capsone Dyon electric charge and fermion fractionalization}
\sjump
\centerline{\capsone in $N=2$ gauge theory}
\bjump\sjump
\centerline{\scaps  Timothy J. Hollowood}
\sjump
\centerline{\sl Department of Physics, University of Wales Swansea,}
\centerline{\sl Singleton Park, Swansea SA2 8PP, U.K.}
\centerline{\tt  t.hollowood@swansea.ac.uk}
\sjump
\bjump\bjump
\ifdoublepage
\vfill
\eject\null\vfill\fi
\centerline{\capsone ABSTRACT}\sjump
A first principles calculation of the quantum corrections to the
electric charge of a dyon in an $N=2$ gauge theory with arbitrary
gauge group is presented. These corrections arise
from the fermion fields via the mechanism of fermion
fractionalization. For a dyon whose magnetic charge is a
non-simple co-root, the correction is a discontinuous function on the moduli
space of vacua and the discontinuities occur precisely on co-dimension
one curves on which the dyon decays. In this way, the complete spectrum of 
dyons at weak coupling is found for a theory with any gauge group. It 
is shown how this spectrum is consistent with the semi-classical 
monodromies.
\sjump\vfill
\eject}
 \vfill

\yespagenumbers\pageno=1
%
%

\chapter{Introduction}

It has been known for a long time that magnetic monopoles can acquire electric
charge. This happens in the presence of a $CP$ violating 
theta term where the electric charge receives a contribution
proportional to the magnetic
charge [\Ref{WIT}]. When there are fermion fields in the theory, it
has also been known for a long time that the electric charge of a monopole can
receive additional contributions from these fields at the quantum
level due to a phenomenon known as fermion fractionalization (see
[\Ref{NS2},\Ref{NPS}] and the review article [\Ref{NS}]). 
At a microscopic level,
the Dirac vacuum of the fermion sector in the background of a monopole
has a non-trivial fermion number and since the fermion carry electric
charge, the vacuum has a non-trivial electric charge which expresses itself in
a contribution to the electric charge of the monopole. The fermions
in effect generate an effective theta angle. The effect is non-trivial
because the induced charge depends on the coordinates of the moduli
space of vacua. We shall calculate this effect in an $N=2$
supersymmetric gauge theory with arbitrary gauge group. The
contribution to the electric charge has the remarkable feature that
for certain magnetic charges it
is discontinuous across the moduli space of vacua. This is shown to be
consistent because the 
discontinuities occur precisely on subspaces on which the dyons
decay and explains the picture of [\Ref{FH1}] which concluded that 
the moduli space of vacua is divided into a number cells separated by
walls on which dyons decay. It is important to
calculate the allowed dyon electric charges because the behaviour of
the theory at strong coupling is governed by the regimes where the
dyons become massless [\Ref{SW1}]. The theory of Seiberg and Witten
generalized to arbitrary gauge groups ([\Ref{KLY},\Ref{AF},\Ref{KLT}] 
for SU($n$),
[\Ref{BL}] for SO($2n$), [\Ref{DS1}] for SO($2n+1$), [\Ref{AS}] for Sp($n$),
[\Ref{DS2}] for $G_2$ and [\Ref{AAG}] for all the other exceptional
cases) makes definite predictions for
the electric charges of the light dyons at strong coupling, hence for
the overall consistency of the theory these charges should match those
at weak coupling. (This assumes that there are paths from weak coupling to
the dyon singularities on which the dyons do not decay---as in the
SU(2) theory [\Ref{SW1}].)

The mass of a BPS state in an $N=2$ supersymmetric gauge theory with
gauge group $G$ and Lie algebra $g$ is given as [\Ref{SW1}]
$$
M=\left\vert \bg\cdot\ba_D(u_j)+\bq\cdot\ba(u_j)\right\vert=\left\vert
Q\cdot A^T\right\vert,
\nfr{MBPS}
where $A=(\ba_D(u_j),\ba(u_j))$ are two $r={\rm rank}(g)$ vector functions 
of the gauge invariant
coordinates $u_j$, $j=1,\ldots,r$, of the moduli space of vacua 
and $Q=(\bg,\bq)$ encodes the magnetic and electric charges
of the state with respect to the unbroken U(1)$^r$
symmetry: the exact relationship requires careful
explanation due to the presence of a theta term and quantum corrections. 
The moduli space of vacua has certain generic features. Firstly there are
{\it singularities\/} of co-dimension two on which a certain BPS state 
is massless defined by
$$
Q\cdot A^T=0.
\efr
At these points the U(1)$^r$ low energy effective action description
breaks down and needs to
be supplemented with fields for the massless states.
The singularities form the boundaries of submanifolds of co-dimension
one on which dyons are kinematically at threshold for decay. If the
singularity corresponds to $Q$ becoming massless then this curve,
or Curve of Marginal Stability (CMS), allows processes of the form
$Q_1\rightarrow Q_2\pm Q$ [\Ref{SW1},\Ref{AFS}]. 
This CMS is defined by the condition:
$$
{\rm arg}\left(Q\cdot A^T\right)={\rm arg}\left(Q_2\cdot A^T\right).
\nfr{ABA}
The functions $\ba_{\rm D}(u_j)$ and $\ba(u_j)$ are not single-valued
functions around a singularity. This implies that associated to each
singularity is a cut---a co-dimension one surface whose boundary is
also the singularity. The positions of these cuts is otherwise
arbitrary. Across the cut $A$ is transformed by a monodromy
transformation $A^T\rightarrow MA^T$. This implies that the charges 
of BPS states are transformed $Q\rightarrow QM^{\pm1}$ (where $M$ acts by
matrix multiplication to the left) as one crosses the cut in either of
the two directions. If a state $Q_1$ is in the spectrum at a particular point
then $Q_1M^{\pm1}$ will be as well, unless the path around the singularity
passes through a CMS on which the dyon decays. 

\chapter{The moduli space and BPS states at weak coupling}

At weak coupling the moduli space has a very simple description since
it is parameterized by the classical Higgs VEV. Up to global
gauge transformations we can take the VEV $\Phi_0$ to be in the Cartan
subalgebra of $g$. So $\Phi_0=\ba\cdot\bH$ which defines the
$r$-dimensional complex
vector $\ba$. However, this does not completely fix the gauge symmetry
since it leaves the freedom to perform discrete gauge transformations
in the Weyl group of $G$. 
This then defines the classical moduli space of vacua:
$$
{\cal M}_{\rm vac}^{\rm cl}\simeq{\Bbb C}^r\big/W(G),
\efr
where $W(G)$ is the Weyl group of $G$. More concretely, we can take 
${\cal M}_{\rm vac}^{\rm cl}$ to be the region for which
${\rm Re}(\ba)$ is in the fundamental Weyl chamber with respect to some
choice of simple roots $\bal_i$:
$$
{\rm Re}(\ba)\cdot\bal_i\geq0,\qquad i=1,2,\ldots,r.
\nfr{RCON}
This also defines a notion of a positive or negative root and we write
$\Phi=\Phi^+\cup\Phi^-$, where $\Phi$ is the root system of the
Lie algebra $g$ and $\Phi^\pm$ are the set of positive and negative roots.
Notice that the wall ${\rm Re}(\ba)\cdot\bal_i=0$ is
identified with itself by $\sigma_i$, the Weyl reflection in $\bal_i$.
As long as $\bal_i\cdot\ba\neq0$, for some $i$, the unbroken gauge group is
abelian U(1)$^r$ generated by the Cartan subalgebra
$\bH$. Classically, therefore, 
the submanifolds where $\bal_i\cdot\ba=0$ correspond to points where
additional gauge bosons become massless. In the quantum theory these
points correspond to regions where the theory becomes strongly coupled
and perturbation theory breaks down. In the neighbourhood of these
singularities the classical moduli space is no longer an appropriate
description. As long as we avoid the neighourhoods of these
singularities the theory is weakly coupled and semi-classical methods
are applicable. As an example, Fig.~1 shows the classical moduli space for
SU(3) where one of the imaginary directions for $\ba$ is suppressed.
The figure shows the singularities and the cuts as the surfaces
identified by the broken arrowed lines.

\midinsert{\bjump
\centerline{
\psfig{figure=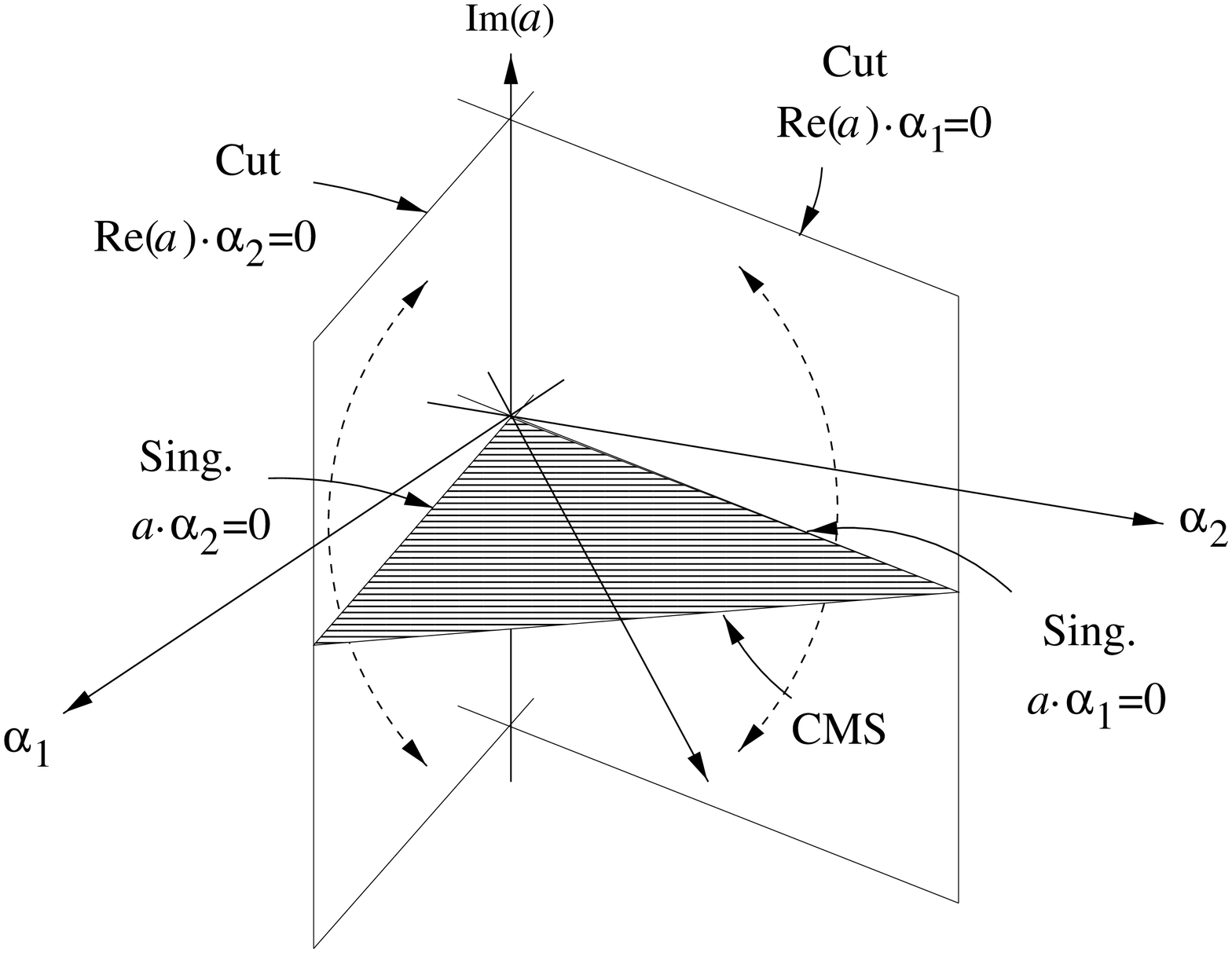,height=7cm}}
\bjump\bjump
\centerline{Figure 1. Classical moduli space for SU(3)}
\sjump
}
\endinsert

Consider the classical theory of monopoles and dyons in an $N=2$
supersymmetric theory. The
mass of a BPS state can be expressed as
$$
M={1\over e}\left\vert Q_E+iQ_M\right\vert,
\efr 
where $Q_E$ and $Q_M$ are given by integrals over a large sphere at
infinity of the inner product in the Lie algebra $g$ 
of the electric and magnetic fields with the Higgs field:
$$
Q_E=\int d\vec S\cdot{\rm Tr}(\vec E\Phi),\qquad
Q_M=\int d\vec S\cdot{\rm Tr}(\vec B\Phi).
\nfr{CDEF}
In unitary gauge all the fields outside the core of dyon
solution are valued in the Cartan subalgebra of $g$. This allows us to
define the vector electric and magnetic charges of the solution:
$$
\bq_{\rm phys}\cdot\bH={1\over e}\int d\vec S\cdot\vec E,\qquad
\bg\cdot\bH={e\over4\pi}\int d\vec S\cdot\vec B.
\nfr{PC}
In the above, $\bq_{\rm phys}$ is the physical vector electric charge
not to be confused with $\bq$ introduced below.
The constants above are included for convenience.
Since on the sphere at spatial infinity the Higgs field is equal
to $\Phi_0=\ba\cdot\bH$ we have
$$
Q_E=e\bq_{\rm phys}\cdot\ba,\qquad Q_M={4\pi\over e}\bg\cdot\ba.
\nfr{CEMC}

The allowed electric and magnetic charges are quantized in the
following way.
The magnetic charge vector $\bg$ has to be a
positive co-root of $g$, i.e. a vector of the form:
$$
\bg=\sum_{i=1}^rn_i\bal_i^\star,
\efr
where $n_i\in{\Bbb Z}\geq0$ and the co-root is defined as
$\bal^\star=\bal/\bal^2$.
The allowed electric charges of the dyon are determined by the
generalized Dirac-Zwanziger-Schwinger (DZS)
quantization condition that requires\note{The
quantization condition actually allows $\bq\in\Lambda_W$, the weight lattice,
however since in the pure gauge theory all the fields 
are adjoint valued only the subset
$\Lambda_R\subset\Lambda_W$ is actually realized.}
$$
\bq\equiv\bq_{\bf phys}-{\theta e\over2\pi}\bg\in \Lambda_R.
\nfr{GDQC}
In the above $\bq$ is not the physical electric charge, rather it is
the vector of eigenvalues of the Noether charges corresponding to
global gauge transformations in the unbroken ${\rm U}(1)^r$ global
symmetry group. It is well-known that in the presence of a theta term
this is not equivalent to the physical electric charge [\Ref{WIT}]. 

Putting this together we see that we can label a BPS
configuration with the two vectors 
$$
Q=(\bg,\bq)\in\left(\Lambda_R^\star,\Lambda_R\right),
\efr
and the mass of the state may be written
$$
M=\left\vert Q\cdot A^T\right\vert,
\nfr{CBPS}
where $\ba_{\rm D}=\tau\ba$ with
$$
\tau={\theta\over2\pi}+{4\pi i\over e^2}.
\efr

Single monopole solutions can be found by embeddings of the
SU(2) monopole solution associated to 
a regular embedding of the Lie algebra $su(2)$ in the Lie algebra of 
the gauge group associated to a positive root $\bal$ and
defined by the three generators $E_{\pm\bal}$ and
$\bal^\star\cdot\bH$. We
denote by $\phi^\bal$ and $\vec A^\bal$ the SU(2) monopole solution
(in the $A_0=0$ gauge) embedded
in the theory with a larger gauge group 
using this $su(2)$ Lie subalgebra. The ansatz for the monopole
solution is [\Ref{AY},\Ref{TJH1}]
$$
\Phi=e^{i\theta_\bal}\left(
\phi^\bal+\hat a\right),\qquad\vec A=\vec A^\bal,
\nfr{ANSATZ}
where the phase angle $\theta_\bal={\rm arg}(\ba\cdot\bal)$ and
$$
\hat a=e^{-i\theta_\bal}\left(\ba-(\ba\cdot\bal^\star)
\bal\right)\cdot\bH.
\efr
The VEV of the SU(2) Higgs field $\phi^\bal$, is fixed to be
$\vert\ba\cdot\bal\vert$, so that $\Phi$ has the correct VEV. 
The solution has magnetic charge vector $\bg=\bal^\star$.
This ansatz is a direct generalization of the ansatz for theories with
a single real Higgs field in [\Ref{BAIS},\Ref{WB1}].

To describe a dyon at leading order in the semi-classical
approximation we note that a
monopole solution generically admits four zero modes
corresponding the position of the centre-of-mass and a periodic U(1)
charge angle. In the leading order semi-classical analysis, momentum
in the periodic direction corresponds to giving the monopole electric
charge along $\bal$. The dyon receives a contribution to the electric charge
from this momentum $\eta$ and from the theta term [\Ref{WIT}]:
$$
\bq_{\rm phys}={\theta
e\over2\pi}\bal^\star+\eta\bal,\qquad{\rm i.e.}\ \bq=\eta\bal.
\nfr{KKLL}
We now demand that the semi-classical wavefunction is invariant under
gauge transformations in the centre of the gauge group. This is
equivalent to the DZS quantization condition \GDQC\ which requires that
$\eta\in{\Bbb Z}$.

Notice that at leading order in the semi-classical approximation
dyons have $\bg\propto\bq$. This means that at leading order it is
straightforward to find the CMS for a particular dyon. A dyon 
whose magnetic charge $\bal^\star$ is a non-simple co-root can decay into a 
number of dyons of magnetic charge $\bgamma^\star$ and $\bd^\star$, 
where $\bgamma$ and $\bd$ are two positive roots of $g$. The CMS for this 
process, which we denote $C_{\bgamma,\bd}$, 
is the submanifold given by \ABA. Defining $\theta_\bal={\rm
arg}(\ba\cdot\bal)$ it is defined by
$$
\theta_\bgamma=\theta_\bd.
\nfr{CMSD}
This surface, along with the walls of the complex Weyl chamber,
divides the classical moduli space into two cells which we
denote $D_{\bgamma,\bd}^\epsilon$ with
$$
\epsilon={\rm sign}\left(\theta_\bgamma-\theta_\bd\right).
\efr
We will find later that quantum effects can give corrections to $\bq$
which are not proportional to $\bg$. These corrections lead to
corrections of the positions of the CMS at higher order, an effect
that do not analyse here.

For a dyon of given magnetic charge $\bal^\star$ we have the following
general description. First of all we need to introduce a set of
rank 2 subspaces of the root system $\Phi$ of $g$. These subspaces of
roots form the set of roots of either an $A_2$, $B_2$ or $G_2$
subalgebra.\note{The latter only occurs in $G_2$ itself, in which case the
subspace is full root space.} Let $\bgamma_i$ and $\bd_i$
be the two positive roots which are the simple roots of the $i^{\rm
th}$ rank 2 subalgebra. Any positive root of the
original algebra in the subspace may be expanded in terms of
$\gamma_i$ and $\bd_i$ with non-negative coefficients. We choose
$\bd_i$ to be the shorter of the two roots: $\bd_i^2\leq\bgamma_i^2$. 

For each of the $L$ 
rank 2 subspace for which $\bal$ is a non-simple root, i.e.
$$
\bal=n_i\bgamma_i+m_i\bd_i,\qquad i=1,\ldots,L,
\efr
for two integers $n_i,m_i\geq1$, there is a CMS $C_{\bgamma_i,\bd_i}$
and two cells $D_{\bgamma_i,\bd_i}^{\epsilon_i}$.
So for these dyons ${\cal M}_{\rm vac}^{\rm cl}$ is divided into 
$2^L$ cells given by the following intersections:
$$
D_{\bgamma_1,\bd_1}^{\epsilon_1}\cap
D_{\bgamma_2,\bd_2}^{\epsilon_1}\cap\cdots\cap
D_{\bgamma_L,\bd_L}^{\epsilon_L},
\nfr{DCELL}
where $\epsilon_i={\rm sign}(\theta_{\bgamma_i}-\theta_{\bd_i})$.
For example, for the case of SU(3), dyons with magnetic charge
$\bal_3^\star=\bal_1^\star+\bal_2^\star$ have a CMS
$C_{\bal_2,\bal_1}$ indicated by the hatched
surface in Fig.~1. This divides
the moduli space into two cells $D^\pm\equiv D^{\pm1}_{\bal_2,\bal_1}$.
Notice that the CMS passes through the two
singularities $\ba\cdot\bal_1=0$ and $\ba\cdot\bal_2=0$, as expected.

\chapter{Quantum corrections}

In the quantum theory, we shall have to re-assess the DZS
quantization condition due to the presence of fermions.
First of all, the electric and magnetic charges in \CEMC\ receive
quantum corrections due to the renormalization of the gauge coupling
and the generation of effective theta terms for the unbroken
U(1)$^r$ symmetry. These effects can be read off from the low energy
effective action. To one-loop
[\Ref{KLY},\Ref{AF},\Ref{KLT},\Ref{BL},\Ref{DS1}]:
$$
\ba_{\rm D}={i\over\pi}\sum_{\bb\in\Phi}\bb(\bb\cdot\ba)\ln\left(
{\bb\cdot\ba\over\Lambda}\right).
\nfr{AD}
Writing $\ba_{\rm D}=\bar\tau\ba$, where
$$
\bar\tau={\bar\theta\over2\pi}+{4\pi i\over\bar{e}^2},
\efr
is a matrix quantity, we find
$$
\bar\theta=-2\sum_{\bb\in\Phi}\left(\bb\otimes\bb\right)
{\rm arg}\left(\bb\cdot\ba\right).
\nfr{ELTP}
The above calculation was done with zero bare theta parameter which
contributes simply a constant term to \ELTP. This contribution can
always be set to zero by performing a U$(1)_R$ transformation which
acts on the Higgs VEV as $\ba\rightarrow e^{i\epsilon}\ba$. 

The form of \ELTP\ shows how 
quantum corrections generate theta-like terms in the effective
action which depend non-trivially on the moduli space coordinate.
We can now read-off the relation between the physical electric charge
and the quantum number $\bq$: 
$$
\bq_{\rm phys}=\bq-
{1\over\pi}\sum_{\bb\in\Phi}\bb\left(\bb\cdot\bg\right)
{\rm arg}\left(\bb\cdot\ba\right),
\nfr{RQQP}
The important point is that the DZS quantization condition still requires
$\bq\in\Lambda_R$. In order to find the 
quantum numbers $\bq$ for the dyons in the theory
we must calculate the physical electric charge $\bq_{\rm phys}$ of the dyons to
one-loop and then extract $\bq$ from \RQQP. 

These one-loop corrections to the electric charge of a dyon are due to the
fermion fields in the theory [\Ref{NS2},\Ref{NPS},\Ref{NS}]. 
The point is that in the background of a topologically charged soliton,
the Dirac vacuum of the fermion field can have a non-trivial fermion
number, a phenomenon know as fermion fractionalization. This was
originally discovered by Jackiw and Rebbi [\Ref{JR}] who showed that
fermion number in the background of topological soliton could be
half-integer. A generalization to arbitrary, irrational values was
obtained by Goldstone and Wilczek [\Ref{GW}]. In the present context, 
the result of this phenomenon is that the fermion number depends
non-trivially on the moduli space coordinate, as recognized in
the context of the SU(2) by Ferrari [\Ref{FF1}]. Since the fermions
carry electric charge this means that the vacuum 
also carries a non-trivial electric charge. In the present
situation this means that 
the fermion fields can contribute to the electric charge of
a dyon. The electric charge of a dyon \KKLL\ is modified to
$$
\bq_{\rm phys}={e\theta\over2\pi}\bal^\star+\eta\bal+\bq_f.
\nfr{QTT}
This changes the quantization rule for $\eta$. 

In the $N=2$ pure gauge theory 
there is a single Dirac fermion which transforms in the
adjoint representation of the gauge group. The interaction of the
fermion field with the dyon is described by the Dirac equation
$$
\left[i\gamma^m D_m-{\rm Re}(\Phi)+i\gamma^5{\rm
Im}(\Phi)\right]\Psi=0.
\nfr{DIRAC}
The Dirac field is adjoint valued and can be expanded in a Cartan-Weyl
basis. We show in the Appendix, in the sector associated to the
root $\bb$, the vacuum has a fermion number
$$
N_\bb={\bb\cdot\bal^\star\over\pi}{\rm arctan}\left({{\rm
Re}\left(e^{-i\theta_\bal}\ba\cdot\bb\right)\over
{\rm Im}\left(e^{-i\theta_\bal}\ba\cdot\bb\right)}\right),
\nfr{NNUM}
where in the above arctan$(x)$ is defined over its principal
range between $-\pi/2$ and $\pi/2$. 

Since the component carries an electric charge
$e\bb$ the 
total electric charge of the Dirac vacuum in the presence of the monopole is
$$
e\bq_f=e\sum_{\bb\in\Phi}\bb N_\bb=
{2e\over\pi}\sum_{\bb\in\Phi^+}\bb\left(\bal^\star\cdot\bb\right)
{\rm arctan}\left({{\rm
Re}\left(e^{-i\theta_\bal}\ba\cdot\bb\right)\over
{\rm Im}\left(e^{-i\theta_\bal}\ba\cdot\bb\right)}\right),
\nfr{ECFF}
where we have used the fact that $N_{-\bb}=-N_\bb$.
Notice that the result \ECFF\ reproduces precisely the functional
dependence on $\ba$ of the effective theta parameter \ELTP.
By comparing \ECFF\ with \RQQP\ we can extract the value of the
quantum number $\bq$:
$$
\bq=\eta'\bal+\sum_{\bb\in\Phi^+}\bb\left(\bal^\star\cdot\bb\right)
{\rm sign}\left(\theta_\bb-\theta_\bal\right),
\nfr{EKK}
where we have used the fact that
$\sum_{\bb\in\Phi^+}(\bal^\star\cdot\bb)\bb\propto\bal$ and have
absorbed a contribution along $\bal$ into a new constant $\eta'$. 
In the above, for a positive root $\bb$ the phase angle $\theta_\bb$
is constrained by \RCON\ to be $-\pi/2\leq\theta_\bb\leq\pi/2$. 
Notice immediately that $\bq$ changes discontinuously whenever the
surfaces where $\theta_\bb=\theta_\bal$ are crossed. These are
precisely the CMS which along with the cuts $\bal_i\cdot\ba=0$
separate ${\cal M}_{\rm 
vac}^{\rm cl}$ into cells where the dyons have different allowed
charges. In fact the result
\NNUM\ and hence \EKK\ is only valid if $\theta_\bb\neq\theta_\bal$,
for some $\bb$, otherwise the Dirac equation has additional
zero modes signaling the fact that, as described in
section 5, the dyon will decay.

In order to determine $\bq$, consider the contribution from a pair of roots 
$\bb$ and $\sigma_\bal(\bb)=\bb-2(\bal^\star\cdot\bb)\bal$, the Weyl 
reflection of $\bb$ in $\bal$. If $\sigma_\bal(\bb)\in\Phi^+$ then
since ${\rm sign}\left(\theta_{\sigma_\bal(\bb)}-\theta_\bal\right)=
{\rm sign}\left(\theta_\bb-\theta_\bal\right)$ and
$\bb\cdot\bal=-\sigma_\bal(\bb)\cdot\bal$,
the contribution from the pair of roots $\bb$ and $\sigma_\bal(\bb)$
is proportional to $\bal$. Therefore the only contributions to $\bq$
which are not proportional to $\bal$ arise when $\sigma_\bal(\bb)\in\Phi^-$.
Accordingly we have
$$
\bq=\eta^{\prime\prime}
\bal+\sum_{\bb\in\Delta}\bb\left(\bal^\star\cdot\bb\right){\rm
sign}\left(\theta_\bb-\theta_\bal\right),
\nfr{DRES}
where we have absorbed an additional contribution along $\bal$ into
a new constant $\eta^{\prime\prime}$. The sum is over the set 
$$
\Delta=\left\{\bb\in\Phi^+\mid\ \bb\neq\bal,\ 
\sigma_\bal(\bb)\in\Phi^-\right\}.
\efr
The generalized Dirac quantization condition \GDQC\ requires that
$\bq$ be a root of the Lie algebra and hence determines 
$\eta^{\prime\prime}$ up to an integer. 

At this point, for purposes of illustration, consider the case of gauge
group SU(3). For dyons with a magnetic charge vector that is a
simple root, $\bal_1^\star$ or $\bal_2^\star$, 
there are no vectors in the set $\Delta$; for instance
$\sigma_{1}(\bal_2)=\bal_3=\bal_1+\bal_2\in\Phi^+$. Hence in these
cases $\bq=n\bal_1$ and $n\bal_2$, $n\in{\Bbb Z}$, respectively.
For a magnetic charge $\bal_3^\star=\bal_1^\star+\bal_2^\star$ 
there are two roots
in $\Delta$. Firstly $\bal_1$, with $\sigma_{3}(\bal_1)=-\bal_2\in\Phi^-$,
and secondly $\bal_2$, with 
$\sigma_{3}(\bal_2)=-\bal_1\in\Phi^-$.
In this case ${\cal M}_{\rm vac}^{\rm cl}$ is divided into two
regions by the CMS $C_{\bal_2,\bal_1}$
In the first cell $D^+$, $\theta_{\bal_2}>\theta_{\bal_1}$,
and so by \DRES\ in this region the $\bal_3^\star$
dyon has 
$\bq=\eta^{\prime\prime}\bal_3+(-\bal_1+\bal_2)/2=n\bal_3-\bal_1$, where
$\eta^{\prime\prime}=n-1/2$ 
and the Dirac quantization condition \GDQC\ implies that
$n\in{\Bbb Z}$. 
In the other cell $D^-$, $\theta_{\bal_2}<\theta_{\bal_1}$,
$\bq=\eta^{\prime\prime}
\bal_3+(\bal_1-\bal_2)/2=n\bal_3+\bal_1$, where
$\eta^{\prime\prime}=n+1/2$ with $n\in{\Bbb Z}$.
So the discontinuity either side of the CMS is $\mp\bal_1$. This is
precisely the spectrum of dyon derived in [\Ref{FH1}] using the
semi-classical monodromies.

Now consider the general case. It is useful to analyse the set
$\Delta$ in terms of the set of $L$ rank 2 root subspaces introduced
at the end of section 3. For each such subspace $\bal$ is a
non-simple root, i.e. 
$$
\bal=n_i\bgamma_i+m_i\bd_i,\qquad n_i,m_i\in{\Bbb
Z}\geq1,
\nfr{LLK} 
and there are contributions to the set $\Delta$. (Weyl
reflections of positive roots in simple roots are always positive
roots again, so there are no contributions to $\Delta$ when 
$\bal=\bgamma_i$ or $\bd_i$.) Let $\bb$ be some positive root in the
subspace so that
$$
\bb=c_i\bgamma_i+d_i\bd_i,\qquad c_i,d_i\in{\Bbb
Z}\geq0. 
\nfr{XXS}
From \LLK\ and \XXS\ one can show that
$$
{\rm sign}\left(\theta_\bb-\theta_\bal\right)={\rm
sign}\left(m_ic_i-n_id_i\right){\rm
sign}\left(\theta_{\bgamma_i}-\theta_{\bd_i}\right),
\efr
and therefore the discontinuity in the charge at 
$\theta_\bb=\theta_\bal$ occurs precisely on the CMS 
$C_{\bgamma_i,\bd_i}$.
So for each rank 2 root subspace for which \LLK\ is true, there exist
contributions $\Delta_i\subset\Delta$ giving a
discontinuity in $\bq$ across
$C_{\bgamma_i,\bd_i}$ which follows from \DRES.
The actual discontinuity can be found explicitly on a case-by-case
basis for the three rank 2 algebras. The are 7 different cases 
(1 in $A_2$, 2 in $B_2$ and 4 in $G_2$) to consider as
indicated in Fig.~2. The contributions to $\bq$ from these subspaces
are calculated below using \DRES, 
up to a contribution along $\bal$ which will be
fixed by the DZS quantization condition.

\midinsert{\bjump
\centerline{
\psfig{figure=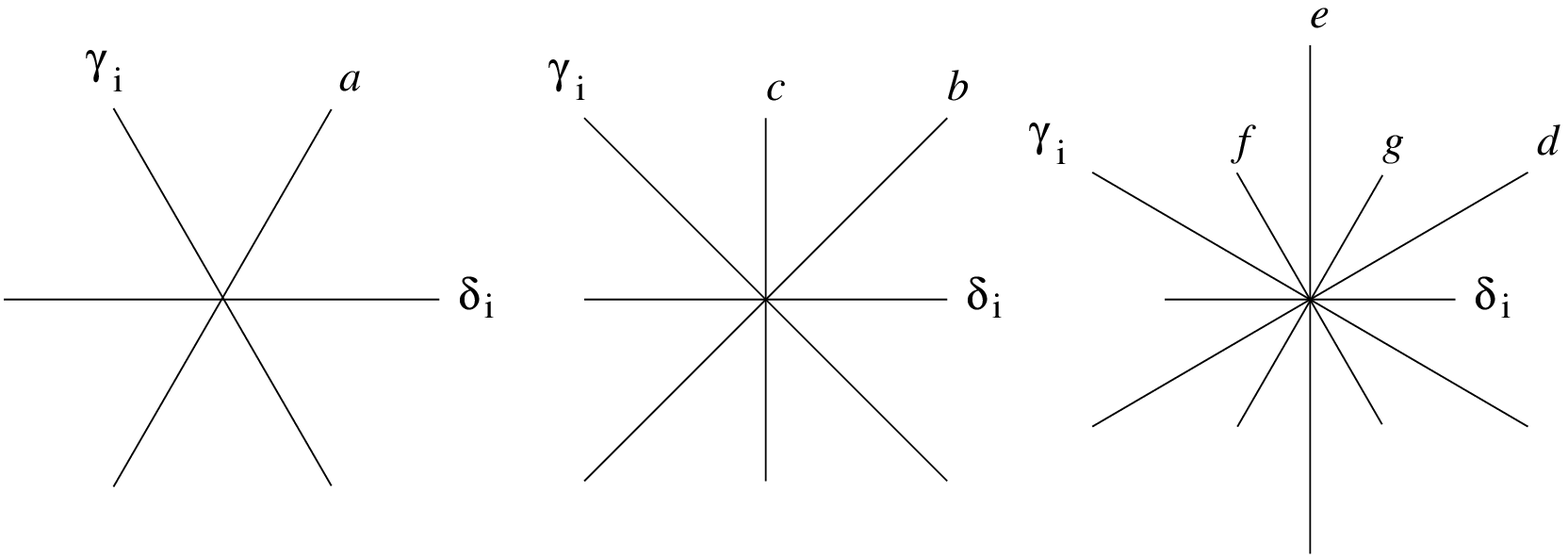,height=4.5cm}}
\bjump\bjump
\centerline{Figure 2. Dynkin diagrams of the rank 2 root subspaces}
\sjump
}
\endinsert

\noindent
($a$) $A_2$ with $\bal=\bgamma_i+\bd_i$ and
$\bal^\star=\bgamma_i^\star+\bd_i^\star$. The set
$\Delta_i=\{\bgamma_i,\bd_i\}$. In the cells 
$D^{\pm1}_{\bgamma_i,\bd_i}$ we find that the contribution to
$\bq$ is $\mp\bd_i$.

\noindent
($b$) $B_2$ with $\bal=\bgamma_i+2\bd_i$ and 
$\bal^\star=\bgamma_i^\star+\bd_i^\star$. 
The set $\Delta_i=\{\bd_i,\bgamma_i+\bd_i\}$. In the cells
$D^{\pm1}_{\bgamma_i,\bd_i}$ we find that the contribution to $\bq$
is $\mp\bd_i$

\noindent
($c$) $B_2$ with $\bal=\bgamma_i+\bd_i$ and 
$\bal^\star=2\bgamma_i^\star+\bd_i^\star$.
The set $\Delta_i=\{\bgamma_i,\bgamma_i+2\bd_i\}$. In the cells
$D^{\pm1}_{\bgamma_i,\bd_i}$ we find that the contribution to $\bq$
is $\mp2\bd_i$

\noindent
($d$) $G_2$ with $\bal=\bgamma_i+3\bd_i$ and
$\bal^\star=\bgamma_i^\star+\bd_i^\star$. The set
$\Delta_i=\{\bgamma_i+2\bd_i,\bd_i\}$. In the cells
$D^{\pm1}_{\bgamma_i,\bd_i}$ we find that the contribution to $\bq$
is $\mp\bd_i$

\noindent
($e$) $G_2$ with $\bal=2\bgamma_i+3\bd_i$ and 
$\bal^\star=2\bgamma_i^\star+\bd_i^\star$. The set
$\Delta_i=\{\bgamma_i,\bgamma_i+\bd_i,\bgamma_i+2\bd_i,\bgamma_i+
3\bd_i\}$. In the cells
$D^{\pm1}_{\bgamma_i,\bd_i}$ we find that the contribution to $\bq$
is $\mp2\bd_i$

\noindent
($f$) $G_2$ with $\bal=\bgamma_i+\bd_i$ and 
$\bal^\star=3\bgamma_i^\star+\bd_i^\star$.
The set
$\Delta_i=\{\bgamma_i,2\bgamma_i+
3\bd_i\}$. In the cells
$D^{\pm1}_{\bgamma_i,\bd_i}$ we find that the contribution to $\bq$
is $\mp3\bd_i$

\noindent
($g$) $G_2$ with $\bal=\bgamma_i+2\bd_i$ and
$\bal^\star=3\bgamma_i^\star+2\bd_i^\star$. The set
$\Delta_i=\{\bgamma_i+\bd_i,\bgamma_i+
3\bd_i,2\bgamma_i+3\bd_i,\bd_i\}$. In the cells
$D^{\pm1}_{\bgamma_i,\bd_i}$ we find that the contribution to $\bq$
is $\mp4\bd_i$

These results can be summarized by a single formula for all cases.
In the cells $D^{\pm1}_{\bgamma_i,\bd_i}$ on either side of the CMS
$C_{\bgamma_i,\bd_i}$ where
$$
\bal=n_i\bgamma_i+m_i\bd_i,\qquad
\bal^\star=N_i\bgamma_i^\star+M_i\bd_i^\star,
\efr
for $n_i,m_i,N_i,M_i\in{\Bbb Z}\geq1$ and $N_i=n_i\bgamma_i^2/\bal_i^2$ and
$M_i=m_i\bd_i^2/\bal^2$,
where $\bd_i$ is the shorter of the two roots $\bgamma_i$ and $\bd_i$,
the contribution to $\bq$ is $\mp(N_i+M_i-1)\bd_i$.
From this we are led to the following general description of the spectrum of
dyons with magnetic charge $\bal^\star$. In each of the cells \DCELL\ the
dyons have
$$
\bq=n\bal-\sum_{i=1}^L(N_i+M_i-1){\rm
sign}\left(\theta_{\bgamma_i}-\theta_{\bd_i}\right)\bd_i,\qquad n\in{\Bbb Z}.
\nfr{BRES}
It follows immediately that the dyons whose magnetic charge is a
simple co-root $\bal_i^\star$ have $\bq=n\bal_i$.

\chapter{Comparison with the semi-classical monodromies}

We have calculated the electric charge quantum by considering the
fermion fractionalization in the background of a monopole. These 
quantum numbers can be calculated independently by following the quantum
numbers of BPS states as one moves around the classical singularities
$\bal_i\cdot\ba=0$ in moduli space [\Ref{FH1}]. The point is that the function
$\ba_D$ is not single-valued as one moves around a singularity due to
the branch-cut of the logarithms from the one-loop contribution in
\AD. A point with coordinate $\ba$ 
on the wall containing the singularity $\bal_i\cdot\ba=0$, 
i.e.~${\rm Re}(\bal_i\cdot\ba)=0$, is identified with a point
$\sigma_{i}(\ba)$. A closed path around the singularity requires
an identification of two such points and one finds that under this 
$$
\left({\ba_{\rm D}\atop\ba}\right)\rightarrow
M_i^{\pm1}
\left({\ba_{\rm D}\atop\ba}\right)=\pmatrix{\sigma_i&
\mp2 \bal_i\otimes\bal_i\cr
0&\sigma_i\cr}\left({\ba_{\rm D}\atop\ba}\right),
\nfr{SCMO}
where $\sigma_i$ is the Weyl reflection in the
simple root $\bal_i$. The signs in \SCMO\
are determined by whether the path circles the singularity in a
positive or negative sense. This is determined by whether the path goes
from a point on the wall with ${\rm Im}(\bal_i\cdot\ba)>0$ to  
one with ${\rm Im}(\bal_i\cdot\ba)<0$, or vice-versa.
Hence in going around the singularity in a positive or negative sense
the quantum numbers $Q=(\bg,\bq)$ of a BPS state are transformed as
$$
Q\rightarrow QM_i^{\pm1},
\efr
respectively,
where $M_i$ acts by matrix multiplication to the left.
To build up a picture of the spectrum one starts with the states
associated to the simple roots with
$Q_i=(\pm\bal_i^\star,n\bal_i)$ which exist throughout the
classical moduli space and then considers taking these states along
various paths around the singularities.

\midinsert{\bjump
\centerline{
\psfig{figure=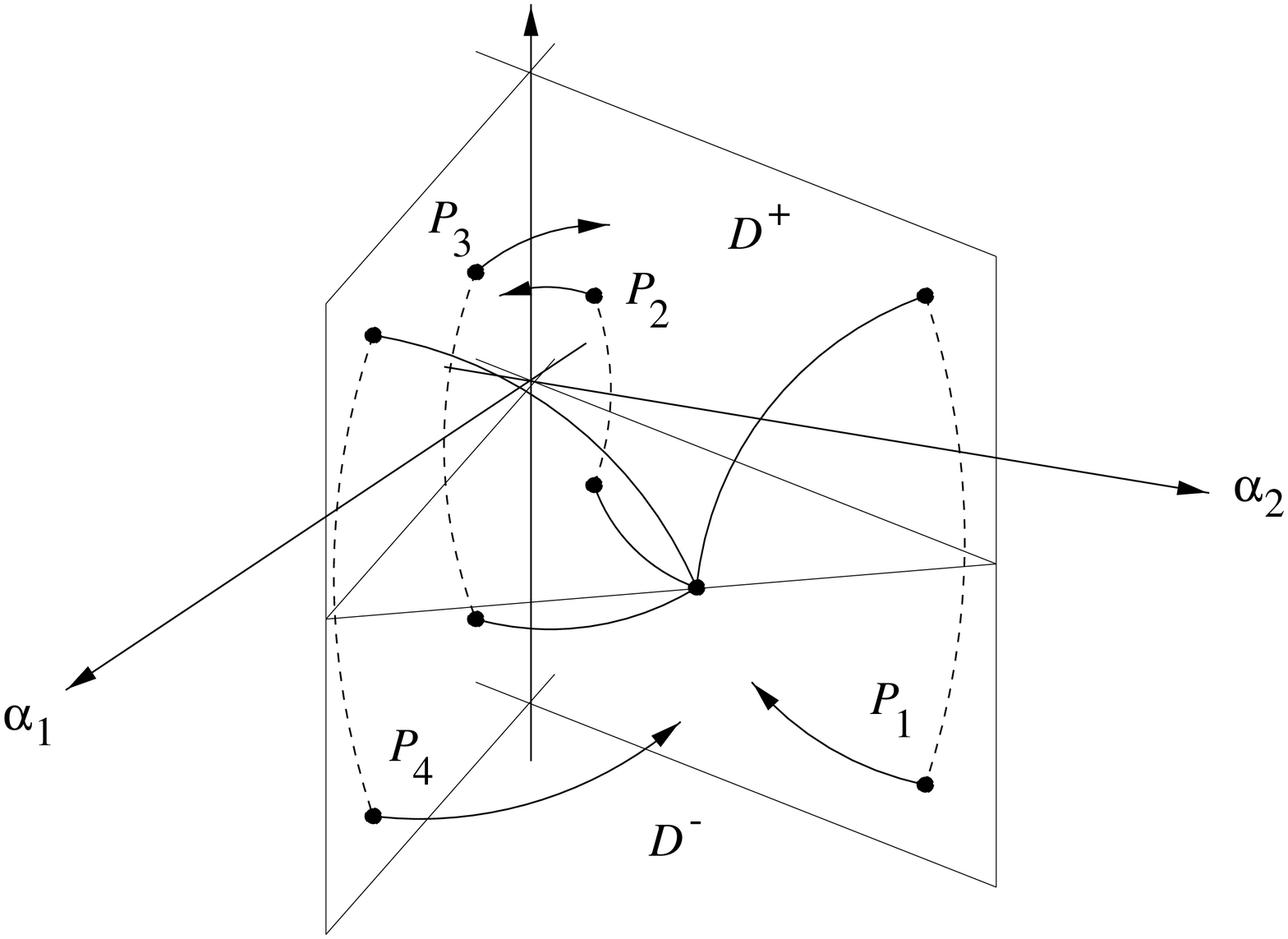,height=7cm}}
\bjump\bjump
\centerline{Figure 3. Paths with non-trivial monodromies for SU(3)}
\sjump
}
\endinsert

For example for the SU(3) case, described in [\Ref{FH1}],
Fig.~2 shows four paths 
$P_i$, $i=1,2,3,4$, which encircle singularities and 
involve the 
monodromy transformations $M_1$, $M_1^{-1}$, $M_2$ and $M_2^{-1}$,
respectively. Starting with the states $Q_1=(\pm\bal_1^\star,n\bal_1)$ and
$Q_2=(\pm\bal_2^\star,n\bal_2)$ one finds that the two cells
$D^\pm$ contains dyons with charges
$$
Q_2M_1^{\mp1}\equiv Q_1M_2^{\pm1},
\efr
respectively. So in $D^\pm$ there are states with
quantum numbers
$$
D^\pm:\qquad\left(\pm\bal_3^\star,n\bal_3\mp\bal_1\right)\equiv\left
(\pm\bal_3^\star,n\bal_3\pm\bal_2\right).
\efr
This matches precisely the result from fermion fractionalization
in \BRES.

This equivalence between the two methods of determining the spectrum
can be extended to any Lie group. The strategy for proving this is to
show that it is true for 
all the rank 2 algebras and then the result for general
groups follows by an inductive argument.
The Dynkin diagrams of all the rank two algebras are illustrated in
Fig.~2, where $\bal_1\equiv\bd_i$ and $\bal_2\equiv\bgamma_i$.
The moduli spaces of all the rank 2 cases are identical
in form to the SU(3) case illustrated in Fig.~1. 
There is a single decay curve defined by
$\theta_{\bal_1}=\theta_{\bal_2}$.
There are 6 additional cases to consider, with $\bal$ equal to the roots
labelled $b,c,\ldots,g$ in Fig.~2.

\noindent
($b$) $B_2$ with $\bal=2\bal_1+\bal_2$. 
A dyon with magnetic charge $(2\bal_1+\bal_2)^\star$ can decay on the CMS to
$(\bal_1^\star)+(\bal_2^\star)$. In the 
cells $D^\pm$ the result \BRES\ gives 
$$
\bq=n(\bal_1+\bal_2)\mp\bal_1,
\nfr{BTT}
respectively.
Since $\sigma_{1}(\bal_2)=2\bal_1+\bal_2$,
these states will be generated by taking  dyons with magnetic charge
$\bal_2^\star$ around
the singularity $\ba\cdot\bal_1=0$. The monodromy gives charges 
$$
Q_2M_1^{\pm1}=
\left((\bal_1+\bal_2)^\star,n(\bal_1+\bal_2)\pm\bal_1\right),
\efr
in $D^\mp$, respectively, which matches \BTT\ exactly.

\noindent
($c$) $B_2$ with $\bal=\bal_1+\bal_2$. A dyon with magnetic charge
$(\bal_1+\bal_2)^\star$ can decay on the CMS to
$2(\bal_1^\star)+(\bal_2^\star)$. In the cells $D^\pm$ the result \BRES\ gives 
$$
\bq=n(\bal_1+\bal_2)\mp2\bal_1,
\nfr{BTO}
respectively.
Since $\sigma_{2}(\bal_1)=\bal_1+\bal_2$, 
these states will be generated by taking a dyon with magnetic charge 
$\bal_1^\star$ around
the singularity $\ba\cdot\bal_2=0$. The monodromy gives charges
$$\eqalign{
Q_1M_2^{\pm1}&=
\left((\bal_1+\bal_2)^\star,n(\bal_1+\bal_2)\pm2\bal_2\right)\cr
&\equiv\left((\bal_1+\bal_2)^\star,n'(\bal_1+\bal_2)\mp2\bal_1\right)\cr}
\efr
in $D^\pm$, respectively, which matches \BTO\ exactly.

\noindent
($d$) $G_2$ with $\bal=3\bal_1+\bal_2$.
A dyon of magnetic charge $(3\bal_1+\bal_2)^\star$ can decay on the
CMS to $(\bal_1^\star)+(\bal_2^\star)$. In the 
cells $D^\pm$ the result \BRES\ gives 
$$
\bq=n(3\bal_1+\bal_2)\mp\bal_1,
\nfr{GTR}
respectively.
Since $\sigma_1(\bal_2)=3\bal_1+\bal_2$,
dyons with magnetic charge $(\bal_1+\bal_2)^\star$ are generated by
taking dyons with magnetic charge $\bal_2$ around the singularity
$\ba\cdot\bal_1$. The monodromies lead to dyons with charges
$$
Q_2M^{\pm1}_1=\left((3\bal_1+\bal_2)^\star,n(3\bal_1+\bal_2)
\pm\bal_1\right),
\efr
in cells $D^\mp$, respectively, which matches \GTR\ precisely.

\noindent
($e$) $G_2$ with $\bal=3\bal_1+2\bal_2$. 
A dyon of magnetic charge $(3\bal_1+2\bal_2)^\star$ can decay to
$(\bal_1^\star)+2(\bal_2^\star)$. In the 
cells $D^\pm$ the result \BRES\ gives 
$$
\bq=n(3\bal_1+2\bal_2)\mp2\bal_1,
\nfr{GTF}
respectively.
Since $\sigma_2\sigma_1(\bal_2)=3\bal_1+2\bal_2$,
dyons with magnetic charge $(3\bal_1+2\bal_2)^\star$ are generated by
taking dyons with magnetic charge $\bal_2$ around the singularity
$\ba\cdot\bal_1$ in a positive (negative) sense and then around the
singularity $\bal\cdot\bal_2=0$ in a positive (negative) sense. 
The monodromies lead to dyons with charges
$$\eqalign{
Q_2M^{\pm1}_1M^{\pm1}_2&=\left((3\bal_1+\bal_2)^\star,n(3\bal_1+\bal_2)
\mp\bal_1\right)M^{\pm1}_2\cr
&=\left((3\bal_1+2\bal_2)^\star,n(3\bal_1+2\bal_2)
\mp(\bal_1+\bal_2)\mp\bal_2\right)\cr
&\equiv\left((3\bal_1+2\bal_2)^\star,n'(3\bal_1+2\bal_2)\mp2\bal_1\right),\cr}
\efr
in cells $D^\pm$, respectively, which matches \GTF\ precisely. It
is important that the monodromy is equal to $M_1^\pm M_2^\pm$ and not
$M_1M_2^{-1}$ or $M_1^{-1}M_2$. These latter monodromy transformations cannot
be realized on the states $Q_2$ because they involve paths that
cross the CMS where the dyons $Q_2M_1$ and $Q_2M_1^{-1}$, respectively, decay.

\noindent
($f$) $G_2$ with $\bal=\bal_1+\bal_2$. A
dyon of magnetic charge $(\bal_1+\bal_2)^\star$ can decay to
$(\bal_1^\star)+3(\bal_2^\star)$. In the 
cells $D^\pm$ the result \BRES\ gives 
$$
\bq=n(\bal_1+\bal_2)\mp3\bal_1,
\nfr{GTO}
respectively.
Since $\sigma_{2}(\bal_1)=\bal_1+\bal_2$,
dyons with magnetic charge $(\bal_1+\bal_2)^\star$ are generated by
taking dyons with magnetic charge $\bal_1$ around the singularity
$\ba\cdot\bal_2$. The monodromies lead to dyons with charges
$$\eqalign{
Q_1M^{\pm1}_2&=\left((\bal_1+\bal_2)^\star,n(\bal_1+\bal_2)
\pm3\bal_2\right)\cr
&\equiv\left((\bal_1+\bal_2)^\star,n'(\bal_1+\bal_2)\mp3\bal_1\right),\cr}
\efr
in cells $D^\pm$, respectively, which matches \GTO\ precisely.

\noindent
($g$) $G_2$ with $\bal=2\bal_1+\bal_2$.
A dyon of magnetic charge $(2\bal_1+\bal_2)^\star$ can decay to
$2(\bal_1^\star)+3(\bal_2^\star)$.  In the 
cells $D^\pm$ the result \BRES\ gives
$$
\bq=n(2\bal_1+\bal_2)\mp4\bal_1,
\nfr{GTT}
respectively.
Since $\sigma_{1}\sigma_{2}(\bal_1)$,
dyons with magnetic charge $(2\bal_1+\bal_2)^\star$ are generated by
taking dyons with magnetic charge $\bal_1$ around the singularity
$\bal\cdot\bal_2=0$ in a positive (negative) sense and then around the
singularity $\bal\cdot\bal_1=0$ in a positive (negative) sense. 
The monodromies lead to dyons with charges
$$\eqalign{
Q_1M^{\pm1}_2M^{\pm1}_1&=\left((\bal_1+\bal_2)^\star,n(\bal_1+\bal_2)
\pm3\bal_2\right)M^{\pm1}_1\cr
&=\left((2\bal_1+\bal_2)^\star,n(2\bal_1+\bal_2)
\pm3(3\bal_1+\bal_2)\pm\bal_1\right)\cr
&\equiv\left((2\bal_1+\bal_2)^\star,n'(2\bal_1+\bal_2)\pm4\bal_1\right),\cr}
\efr
present in cells $D^{\mp}$, respectively, which matches \GTT\ precisely.

Fortunately it is only a little bit harder to establish the exact relation
between the semi-classical monodromies and the result \BRES\ for all
the gauge groups of rank $>2$. The reason is that it is not necessary
to prove the result case-by-case, since the result can be proved by induction
assuming that it is true for the rank 2 cases, a fact that has
been established above. It is useful at this stage to
consider in more detail the relation between the Weyl group and the
semi-classical monodromies. The Weyl reflections in the simple roots
$\sigma_{j}$ are associated to the pair of monodromy
transformations $M_j^{\pm1}$. In fact the semi-classical monodromies
generate a representation of the Brieskorn Braid Group [\Ref{AVGL}]. To
each element $\sigma$ of the Weyl group $W$, which can be written
$\sigma=\sigma_{a_1}\cdots\sigma_{a_p}$, there exist the following
elements of the Braid group:
$M=M_{a_1}^{\epsilon_1}\cdots M_{a_p}^{\epsilon_p}$, where $\epsilon_i=\pm1$.
These elements all have the form
$$
M=\pmatrix{\sigma&A\cr 0&\sigma\cr},
\nfr{BGE}
for some matrix $A$ depending on the $\epsilon_i$.

Now consider the spectrum of dyons of magnetic charge
$\bal^\star=N_i\bgamma_i^\star+M_i\bd_i^\star$ 
either side of the CMS $C_{\bgamma_i,\bd_i}$. There always exists an element
$\sigma\in W$ such that $\bal_j=\sigma(\bgamma_i)$ and
$\bal_k=\sigma(\bd_i)$, where $\bal_j$ and $\bal_k$ are two simple roots
(with some choice of labelling)
which form either an $A_2$ or $B_2$ type sub-root space of $\Phi$. The
choice of $\sigma$, $\bal_j$ and $\bal_k$ are not unique, but this
will not affect the result. In some cell 
$$
{\cal D}=D^{\epsilon_1}_{\bgamma_1,\bd_1}\cap\cdots\cap
D^{\epsilon_{i-1}}_{\bgamma_{i-1},\bd_{i-1}}\cap 
D^{\epsilon_{i+1}}_{\bgamma_{i+1},\bd_{i+1}}
\cap\cdots\cap D^{\epsilon_L}_{\bgamma_L,\bd_L}
\efr
which straddles $C_{\bgamma_i,\bd_i}$ the dyons of magnetic charge
$\bgamma_i^\star$ and $\bd_i^\star$ have charges
$$
Q_{\bgamma_i}=Q_jM,\qquad Q_{\bd_i}=Q_kM,
\efr
where $M$ is an element of the Braid group of the form \BGE\ for some
$A$ which depends on $\cal D$, i.e.~on the $\epsilon_l$ for $l\neq i$.
To work out the charges of $\bal^\star$ we can map the
problem back to the rank 2 subspace defined by $\bal_j$ and
$\bal_k$. In particular, 
$\sigma(\bal^\star)=N_i\bal_j^\star+M_i\bal_k^\star$ and
$\sigma(\bal)=n_i\bal_j+m_i\bal_k$. We have previously 
established in this section (by identifying $\bal_j$ with $\bal_2$ and
$\bal_k$ with $\bal_1$) 
that in the regions $D_{\bal_j,\bal_k}^{\pm1}$ there
are dyons of charge
$$
Q^\pm=\left(N_i\bal_j^\star+M_i\bal_k^\star,
n(n_i\bal_j+m_i\bal_k)\mp(N_i+M_i-1)\bal_k\right),\qquad n\in{\Bbb Z}.
\efr
By following a path with monodromy $M$, which does not pass through a
CMS on which the dyons decay, these dyons give dyons of charge $Q^\pm M$ in
$D_{\bgamma_i,\bd_i}^{\pm1}\cap{\cal D}$. Explicitly
$$\eqalign{
Q^\pm
M&=\left(\bal^\star,n\bal\mp(N_i+M_i-1)
\sigma^T(\bal_k)+A^T\sigma(\bal^\star)\right)\cr
&=\left(\bal^\star,n\bal+\bb\mp(N_i+M_i-1)\bd_i\right),\cr}
\nfr{HGG}
where $\bb=A^T\sigma(\bal^\star)$ 
is some root which depends on $\cal D$. 
In \HGG\ we have used the expression for $M$ in \BGE\
and the fact that
$\sigma^T=\sigma^{-1}$ implying $\bgamma_i=\sigma^T(\bal_j)$ and 
$\bd_i=\sigma^T(\bal_k)$. Following this argument for each of the
cells in \DCELL\ for $i=1,\ldots,L$, one deduces the general result
\BRES. 

\chapter{Dyon decays}

For the consistency of the spectrum that we have deduced in \BRES\ it
is important that the dyon with magnetic charge $\bal^\star$ decays on
the CMS $C_{\bgamma_i,\bd_i}$, $i=1,\ldots,L$. In certain cases, this
fact can also be established in the semi-classical approximation [\Ref{TJH1}]. 

At a generic point in ${\cal M}_{\rm vac}^{\rm cl}$, a
monopole solution has four bosonic zero modes, corresponding to the
four collective coordinates which specify the
centre-of-mass and the U(1) charge angle
[\Ref{AY},\Ref{TJH1},\Ref{FH2}]. The moduli space of the collective
coordinates has the form $M={\Bbb R}^3\times S^1$. In semi-classical
quantization the states of the monopole are described by a 
supersymmetric quantum mechanics on the moduli space
of collective coordinates [\Ref{G1}]. This yields the usual tower of
dyon states,
where the electric charge corresponds to momentum in the compact
direction. These states transform in a hupermultiplet representation
of $N=2$ supersymmetry. On the CMS $C_{\bgamma_i,\bd_i}$, where 
$\bal^\star=N_i\bgamma_i^\star+M_i\bd_i^\star$, there are $4(N_i+M_i-1)$
additional zero modes. This can be deduced from the index theory
calculation of Weinberg [\Ref{WB1}]. 
The space of collective coordinates 
enlarges discontinuously to space which has the form [\Ref{LWY1}]:
$$
{\Bbb R}^3\times S^1\rightarrow 
M'={\Bbb R}^3\times{{\Bbb R}\times M_0\over{\cal D}},
\efr
where $\cal D$ is a certain discrete group, the factor $\Bbb R$ is associated
with the overall charge angle and $M_0$, the centred moduli space, 
is a hyper-kahler manifold of dimension
$4(N_i+M_i-1)$. The various different cases described previously give the
possible dimensions of $M_0$ as 4,8,12 or 16. In
the case when $\bal^\star=\bgamma_i^\star+\bd_i^\star$, 
$M_0$ has dimension 4 and
is a Euclidean Taub-NUT space [\Ref{LWY1},\Ref{GL}].

The question of whether the dyon with magnetic charge $\bal^\star$
decays on $C_{\bgamma_i,\bd_i}$ can be answered by enquiring whether
or not there is a threshold bound-state in the supersymmetric quantum
mechanics on $M'$ [\Ref{TJH1}]. Such bound-states require a normalizable
holomorphic harmonic form (or spinor) on $M_0$. For the case when
$M_0$ is the Euclidean Taub-NUT space it is known that there is no
such form and so there are no bound-states and the dyons decay. The
description of the spectrum that we have established 
strongly suggests that no such threshold
bound-states exists for the higher-dimensional cases as well.

\chapter{Discussion}

We have constructed the dyon spectrum of $N=2$ gauge theories at weak
coupling for theories with an arbitrary gauge group. The spectrum of
states that arise is consistent with the semi-classical monodromies. The
effect that allows us to determine the charges of the dyon states
involves fermion fractionalization which is a one-loop effect. It is
known that for these BPS states there are no higher perturbative
corrections to the electric charge. In order words the vector quantum
numbers $\bq$ that we have derived are not subject to any additional
corrections. These states can then be followed into the strong
coupling regime as long as one does not cross a CMS on which they
decay. These states should then be responsible for the singularities
that occur at strong coupling. It would be interesting to investigate
whether the dyons present at weak coupling can account for all the
strong coupling singularities as checked for the SU($N$) theory in
[\Ref{FH1}]. 

\bjump
I would like to thank PPARC for an Advanced Fellowship.

\appendix{}

In this appendix we calculate the fractional fermion number of the
vacuum in the background of a monopole. The monopole solution is given
via the embedding in \ANSATZ.

The Dirac equation for the fermion fields in the background of the
monopole solution is
$$
\left[i\gamma^mD_m-{\rm Re}(\Phi)+i\gamma^5{\rm
Im}(\Phi)\right]\Psi=0.
\efr
The background is time-independent and so we look for stationary
solutions $\Psi=e^{iEt}\psi$:
$$
\left[i\gamma^0\gamma^iD_i-\gamma^0{\rm Re}(\Phi)+i\gamma^0\gamma^5
{\rm Im}(\Phi)\right]\psi=E\psi.
\nfr{HEQ}
In order to write this equation in a recognizable form we first
perform a U$(1)_R$ transformation $\Phi\rightarrow e^{-i\theta_\bal}\Phi$.
With a suitable basis for the gamma matrices, the Hamiltonian
equation \HEQ\ now takes the form
$$
H\psi=\pmatrix{{\rm Im}(\hat a)&D\cr D^\dagger&-{\rm Im}(\hat
a)\cr}\psi=E\psi,
\nfr{HDIR}
where 
$$
D=i\sigma_iD_i+i\phi^\bal+i{\rm Re}(\hat a),\qquad
D^\dagger=i\sigma_iD_i-i\phi^\bal-i{\rm Re}(\hat a).
\efr

The fermion number of the vacuum is related to the spectral asymmetry
of the Dirac Hamiltonian $H$. Writing,
$$
H=H_0+{\rm Im}(\hat a)\Gamma^c,\qquad H_0=\pmatrix{0&D\cr D^\dagger&0\cr},
\qquad\Gamma^c=\pmatrix{1&0\cr 0&-1\cr},
\efr
the Hamiltonian $H_0$ anti-commutes with the charge conjugation matrix
$\Gamma^c$. This means that for $H_0$ there is a
precise mapping between states of positive and negative
energy. $H_0$ is precisely the Dirac operator consider by Weinberg for
the theory with a single real Higgs field [\Ref{WB1}].
$H$ on the other hand does not admit a conjugation symmetry. Even so there is
a mapping between positive and negative energy eigenvectors. If
$\psi$ has positive energy $E$ then $[H,\Gamma^c]\psi$ is an
eigenvector with energy $-E$. However, since this mapping involves a
derivative it does not guarantee that the densities of the positive
and negative
energy modes of the continuum part of the spectrum are identical. 
Hence in general
the operator $H$ has a non-trivial spectral asymmetry which is defined
formally as
$$
\eta_H={\rm Tr}\left({\rm sign}(H)\right)={2\over\pi}
\int_0^\infty dz\,{\rm Tr}\left({H\over H^2+z^2}\right),
\efr
where the trace is over the whole Hilbert space and the second
expression involves an integral representation of the sign
function. The fermion number of the vacuum is given by [\Ref{NS}]
$$
N=-\half\eta_H=-{1\over\pi}\int_0^\infty dz\,{\rm Tr}
\left({H\over H^2+z^2}\right)
\efr
This can be written as [\Ref{NS}]:
$$
N={1\over\pi}\int_0^\infty dz\,{\rm Tr}\left({{\rm Im}(\hat a)\over
D^\dagger D+{\rm Im}(\hat a)^2+z^2}-   
{{\rm Im}(\hat a)\over DD^\dagger+{\rm Im}(\hat a)^2+z^2}\right).
\efr
A trace of the kind above was calculated by Weinberg [\Ref{WB1}] by an
adaptation of the Callias index theorem [\Ref{CAL}]. It turns out that
the only non-zero contribution comes from a surface term and hence
only involves the fields at spatial infinity. The relevant result for
the trace is
$$
\sum_{\beta\in\Phi}{{\rm Re}\left(e^{-i\theta_\alpha}\ba\cdot\bb\right)
{\rm Im}\left(e^{-i\theta_\alpha}\ba\cdot\bb\right)\bb\cdot\bal^\star\over
\left(\vert\ba\cdot\bb\vert^2+z^2\right)^{1/2}\left(
{\rm Im}\left(e^{-i\theta_\alpha}\ba\cdot\bb\right)^2+z^2\right)}.
\nfr{TRAC}
The fact that the integrand separates into a sum over all the roots in
the Lie algebra allows us to interpret the result as a sum over the
fractional fermion numbers $N_\bb$ associated to the expansion of the
fermion field in a Cartan-Weyl basis. By evaluating the integral, we
have for each mode
$$
N_\bb={\bb\cdot\bal^\star\over\pi}{\rm arctan}\left({{\rm
Re}\left(e^{-i\theta_\alpha}\ba\cdot\bb\right)\over
{\rm Im}\left(e^{-i\theta_\alpha}\ba\cdot\bb\right)}\right).
\efr

\references

\beginref
\Rref{NS2}{A.J. Niemi and G.W. Semenoff, Phys. Rev. {\bf D30} (1984)
809}
\Rref{NPS}{A.J. Niemi, M.B. Paranjape and G.W. Semenoff,
Phys. Rev. Lett. {\bf 53} (1984) 515}
\Rref{TJH1}{T.J. Hollowood, {\tt hep-th/9611106}}
\Rref{WIT}{E. Witten, Phys. Lett. {\bf B86} (1979) 283}
\Rref{SW1}{N. Seiberg and E. Witten, Nucl. Phys. {\bf B426} (1994) 19,
{\tt hep-th/9407087}}
\Rref{LWY1}{K. Lee, E.J. Weinberg and P. Yi, 
Phys. Lett. {\bf B376} (1996) 97, {\tt hep-th/9601097}; 
Phys. Rev. {\bf D54} (1996) 1633, {\tt hep-th/9602167}}
\Rref{GL}{J.P. Gauntlett and D. A. Lowe, 
Nucl. Phys. {\bf B472} (1996) 194, {\tt hep-th/9601085}}
\Rref{FF1}{F. Ferrari, Phys. Rev. Lett. {\bf78} (1997) 795, {\tt
hep-th/9609101}} 
\Rref{AFS}{P.C. Argyres, A.E. Faraggi and A.D. Shapere, `Future
perspectives in string theory', Los Angeles (1995) 1, 
{\tt hep-th/9505190}}
\Rref{WB1}{E.J. Weinberg, Nucl. Phys. {\bf B167} (1980) 500;  
Nucl. Phys. {\bf B203} (1982) 445}
\Rref{BAIS}{F.A. Bais, Phys. Rev. {\bf D18} (1978) 1206}
\Rref{AY}{O. Aharony and S. Yankielowicz, Nucl.Phys. {\bf B473} (1996)
93, {\tt hep-th/9601011}} 
\Rref{NS}{A.J. Niemi and G.W. Semenoff, Phys. Rep. {\bf135} (1986) 99}
\Rref{FH1}{C. Fraser and T.J. Hollowood, Nucl. 
Phys. {\bf B490} (1997) 217, {\tt hep-th/9610142}}
\Rref{KLY}{A. Klemm, W.Lerche and S. Yankielowicz, Phys. Lett. {\bf
B344} (1995) 169, {\tt hep-th/9411048}}
\Rref{AF}{P.C. Argyres and A.E. Faraggi, Phys. Rev. Lett. {\bf74}
(1995) 3931, {\tt hep-th/9411057}}
\Rref{DS1}{U.H. Danielsson and B. Sundborg, Phys. Lett. {\bf B358}
(1995) 273, {\tt hep-th/9504102}}
\Rref{BL}{A. Brandhuber and K. Landsteiner, Phys. Lett. {\bf B358}
(1995) 73, {\tt hep-th/9507008}}
\Rref{KLT}{A. Klemm, W. Lerche and S. Theisen, Int. J. Mod. Phys. {\bf
A11} (1996) 1929, {\tt hep-th/9505150}}
\Rref{AS}{P.C. Argyres and A.D. Shapere, Nucl. Phys. {\bf B461} (1996)
437, {\tt hep-th/9509175}} 
\Rref{AAG}{M.R. Abolhasani, M. Alishahiha and A.M. Ghezelbash,
Nucl. Phys. {\bf B480} (1996) 279, {\tt hep-th/9606043}}
\Rref{DS2}{U.H. Danielsson and B. Sundborg, Phys. Lett. {\bf B370}
(1996) 83, {\tt hep-th/9511180}}
\Rref{AFS}{P.C. Argyres, A.E. Faraggi and A.D. Shapere, {\tt hep-th/9505190}}
\Rref{AY}{O. Aharony and S. Yankielowicz, Nucl. Phys. {\bf B473} (1996)
93, {\tt hep-th/9601011}}
\Rref{JR}{R. Jackiw and C. Rebbi, Phys. Rev. {\bf D13} (1976) 3398}
\Rref{GW}{J. Goldstone and F. Wilczek, Phys. Rev. Lett {\bf47} (1981) 986}
\Rref{AVGL}{V.I. Arnol'd, V.A. Vasil'ev, V.V. Goryunov and O.V. Lyashko,
`Singularity Theory I, Dynamical Systems VI', Encyclopaedia of
Mathematical Sciences, Vol 6, Springer-Verlag 1993.}
\Rref{CAL}{C.J. Callias, Commun. Math. Phys. {\bf62} (1978) 213}
\Rref{FH2}{C. Fraser and T.J. Hollowood, {\sl to appear in\/}: Phys. 
Lett. {\bf B}, {\tt hep-th/9704011}}
\Rref{G1}{J. Gauntlett, Nucl. Phys. {\bf B411} (1994) 443}
\endref

\ciao

\appendix{B}

Following Weinberg [\Ref{WB1}], the index
$$
{\cal N}=2\lim_{z\rightarrow0}\left[
{\rm Tr}\left({z^2\over D^\dagger D+{\rm Im}(\hat a)^2+z^2}-   
{z^2\over DD^\dagger+{\rm Im}(\hat
a)^2+z^2}\right)\right].
\efr
counts the number of bosonic zero modes of the monopole solution.
From \TRAC\ we have
$$
{\cal N}=2\lim_{z\rightarrow0}\left[\sum_{\beta\in\Phi}{z^2{\rm
Re}\left(e^{-i\theta_\alpha}\ba\cdot\bb\right)\bb\cdot\bal^\star\over 
\left(\vert\ba\cdot\bb\vert^2+z^2\right)^{1/2}\left(
{\rm
Im}\left(e^{-i\theta_\alpha}\ba\cdot\bb\right)^2+z^2\right)}\right]
=4\sum_{\beta\in\Theta}\bb\cdot\bal^\star,
\nfr{RINT}
where the sum is over $\Theta$ the set of positive roots with
$\theta_\bal-\theta_\bb=0$. At a generic point in the moduli space,
$\Theta=\{\bal\}$ only, giving 4 zero modes corresponding to
the centre-of-mass and U(1) charge angle collective coordinates. On a
CMS $C_{\bgamma_i,\bd_i}$ there are additonal roots in $\Theta$
corresponding to the positive roots of the associated rank 2 root
subspace. Following Weiberg, in this case \RINT\ gives
$$
{\cal N}=4(N_i+M_i).
\efr
On intersections of CMS's there will be the appropriate combination of
zero modes.